\documentclass[apj]{emulateapj}

\def\ltsima{$\; \buildrel < \over \sim \;$}
\def\simlt{\lower.5ex\hbox{\ltsima}}
\def\gtsima{$\; \buildrel > \over \sim \;$}
\def\simgt{\lower.5ex\hbox{\gtsima}}
\def\kms{{\rm\,km\,s^{-1}}}

\def\kpc{{\rm\,kpc}}

\def\msun{{\rm\,M_\odot}}
\def\lsun{{\rm\,L_\odot}}

\def\pc{{\rm\,pc}}

\newcommand{\fmmm}[1]{\mbox{$#1$}}

\newcommand{\mcnd}{\mbox{\fmmm{'}\hskip-0.3em .}}

\def\deg{^\circ}

\def\s{\ifmmode \widetilde \else \~\fi}
\def\={\overline}

\def\spose#1{\hbox to 0pt{#1\hss}}

\def\lta{\mathrel{\spose{\lower 3pt\hbox{$\mathchar"218$}}
     \raise 2.0pt\hbox{$\mathchar"13C$}}}
\def\gta{\mathrel{\spose{\lower 3pt\hbox{$\mathchar"218$}}
     \raise 2.0pt\hbox{$\mathchar"13E$}}}
\def\Dt{\spose{\raise 1.5ex\hbox{\hskip3pt$\mathchar"201$}}}    
\def\dt{\spose{\raise 1.0ex\hbox{\hskip2pt$\mathchar"201$}}}    

\def\dotsfill{\leaders\hbox to 1em{\hss.\hss}\hfill}
\def\sun{\odot}

\def\Gyr{{\rm\,Gyr}}
\def\FeH{{\rm[Fe/H]}}

\shorttitle{Structural properties of faint MW satellites}
\shortauthors{N. F. Martin, J. T. A. de Jong \& H.-W. Rix}

\begin{document}


\title{A comprehensive Maximum Likelihood analysis of the structural properties of faint Milky Way satellites}


\author{Nicolas F. Martin$^1$, Jelte T. A. de Jong$^1$ \& Hans-Walter Rix$^1$}
\email{martin,dejong,rix@mpia.de}

\altaffiltext{1}{Max-Planck-Institut f\"ur Astronomie, K\"onigstuhl 17, D-69117 Heidelberg, Germany}

\begin{abstract}
We derive the structural parameters of the recently discovered very low luminosity Milky Way satellites through a Maximum Likelihood algorithm applied to SDSS data. For each satellite, even when only a few tens of stars are available down to the SDSS flux limit, the algorithm yields robust estimates and errors for the centroid, position angle, ellipticity, exponential half-light radius and number of member stars (within the SDSS). This latter parameter is then used in conjunction with stellar population models of the satellites to derive their absolute magnitudes and stellar masses, accounting for `color-magnitude diagram shot-noise'. Most parameters are in good agreement with previous determinations but we now properly account for parameter covariances. However, we find that faint satellites are somewhat more elliptical than initially found and ascribe that to the previous use of smoothed maps which can be dominated by the smoothing (round) kernel. As a result, the faintest half of the Milky Way dwarf galaxies ($M_V\gta-7.5$) is significantly ($4\sigma$) flatter ($<\epsilon>=0.47\pm0.03$) than its brightest half ($M_V\lta-7.5$, $<\epsilon>=0.32\pm0.02$). From our best models, we also investigate whether the seemingly distorted shape of the satellites, often taken to be a sign of tidal distortion, can be quantified. We find that, except for tentative evidence of distortion in Canes Venatici~I and Ursa Major~II, these can be completely accounted for by Poisson scatter in the sparsely sampled systems. We consider three scenarios that could explain the rather elongated shape of faint satellites: rotation supported systems, stars following the shape of more triaxial dark matter subhalos, or elongation due to tidal interaction with the Milky Way. Although none of these is entirely satisfactory, the last one appears the least problematic, but obviously warrants much deeper observations to track evidence of such tidal interaction.
\end{abstract}

\keywords{Local Group --- galaxies: dwarf}

\section{Introduction}
Owing to new sky surveys, the number of known dwarf galaxies or dwarf galaxy candidates that reside within the Local Group has doubled over the last very few years \citep{zucker04,willman05a,willman05b,belokurov06c,martin06b,sakamoto06,zucker06b,zucker06a,belokurov07a,ibata07,majewski07,walsh07,zucker07,irwin08,mcconnachie08}. These new discoveries have extended the realm of galaxies to objects $\sim100$ times fainter than were known before. This enlarged population of extremely dim satellites is bringing us closer to an (at least) partial solution to the apparent overproduction of dark matter sub-halos in cosmological simulations of Milky Way (MW) and M31-like systems \citep{koposov08,penarrubia08a,simon07,strigari07}. This is also supported by the spectroscopic surveys of these objects and measurement of their velocity dispersions that require them to be among the most dark matter dominated objects known to date \citep{kleyna05,munoz06b,martin07a,simon07}.

Although some satellite candidates are still being searched for and proposed \citep[e.g.][]{liu08}, most MW satellites within the reach of the Sloan Digital Sky Survey (SDSS) have probably been discovered \citep{koposov08}. More and more effort is now spent on trying to understand whether these systems reflect the faint end of the dwarf galaxy luminosity function, whether they are disrupted versions of their brighter siblings or whether they are entirely different objects \citep{belokurov07a,gilmore07,penarrubia08b}. But as these studies are based on the observed properties of these systems, it is important to quantify their structural parameters or luminosities rigorously, given their often low number of stars (a few tens to a few hundreds down to the SDSS flux limit).

As a simple comparison, both Bo\"otes~I (BooI) and Draco (Dra) are systems containing mainly (if not only) metal-poor and old stars, at similar distances from us ($\sim75$ vs. $\sim60\kpc$), yet their difference in luminosity ($M_V=-8.8$ vs. $M_V\simeq-6.0$) means that the number of stars observable in BooI is roughly 10\% that of Draco under similar observing conditions. Nevertheless, for the faint satellites, parameters are usually derived with the same techniques that are used for their brighter counterparts. In particular, the structural parameters are mainly derived by assuming spherical symmetry or through the intensity-weighted second moments technique (see e.g. \citealt{mcconnachie06b} for a description). But in faint galaxies, this technique requires some smoothing to produce maps that are not dominated by shot-noise and on which the method can be applied reliably. The impact of any such smoothing scheme (pixel size, smoothing kernel size, threshold over which pixels are used) on the recovered parameters has not been investigated extensively.

In addition, the `total magnitude' of extremely faint stellar systems is conceptually not well defined since this parameter is usually determined from the luminosity of individual stars. This is perfectly valid when thousands of stars provide a good sampling of the color-magnitude diagram (CMD), but as the CMD becomes more sparsely populated in faint satellites, the recovered luminosity can be affected by the stellar evolution state of individual stars, in particular how many of them are close to the tip of the red giant branch. In Willman~1 (Wil1) single stars could have a strong  but transient impact on the luminosity of the system, currently estimated at $\sim1000\lsun$. Finally, the extent to which the seemingly distorted shapes of most of these new systems [e.g. Bo\"otes~I (BooI, \citealt{belokurov06c}), CVnI (\citealt{zucker06a,martin08a}), Coma Berenices (Com, \citealt{belokurov07a}), Hercules (Her, \citealt{coleman07}), Ursa Major~II (UMaII, \citealt{zucker06b}) and Wil1 (\citealt{willman06})] are intrinsic or a reflection of the few stars available to map them also warrants close scrutiny.

Therefore, we set out to re-derive the structural parameters (centroid, half-light radius, ellipticity, position angle) and luminosities of the newly discovered MW satellites through a Maximum Likelihood (ML) algorithm. This algorithm has the advantage of directly fitting the best stellar density model to the data --- the positions of individual stars --- without any need for smoothing. Moreover, one of its outputs is the actual number of stars present in each system, down to the SDSS depth, which we can then use to determine its expected magnitude. This latter value then accounts for the scatter that is induced by the individual evolution of member stars. Finally, we use the best fit structural models to test whether shot-noise alone can explain the seemingly irregular shape of smoothed stellar density maps, or whether they deviate enough from these best models to require clumpiness in their underlying distribution.

This paper is organized as follows: \S~2 describes the Maximum Likelihood algorithm and presents its results applied to SDSS data. We also analyze discrepancies with previous estimates of the ellipticity of some systems. In \S~3 we study the impact of the sparseness of the systems on their absolute magnitude while the search for deviations from the best model is explained in \S~4. Finally, we use \S~5 for a discussion of the ensuing homogeneous dataset of morphological parameters and consider possible scenarios that could explain the higher characteristic ellipticity found in the faintest half of the Milky Way dwarf galaxies. \S~6 concludes this paper.

\section{Structural properties}
\subsection{Maximum Likelihood parameter estimates}
While the approach of estimating structural parameters for stellar systems directly from a star catalog  via a Maximum Likelihood fit is in principle established (e.g. \citealt{kleyna98,westfall06}), it has not been used very widely. Therefore, we recapitulate the approach here and spell out our particular implementation. We start by presuming that the observables, in our case the positions of stars in the SDSS, have been drawn from a (spatial) model distribution that is described by $j$ model parameters, $p_1,p_2,\dots,p_j$. The goal of the well-known ML technique (e.g. \citealt{lupton93}) is to find the set of parameters $(\hat{p}_1,\hat{p}_2,\dots,\hat{p}_j)$ for which the observations become most likely. This means maximizing the $\mathcal{L}$ function defined as:

\begin{equation}
\label{Leq}
\mathcal{L}(p_1,p_2,\dots,p_j) = \prod_i \ell_i(p_1,p_2,\dots,p_j)
\end{equation}

\noindent where $\ell_i(p_1,p_2,\dots,p_j)$ is the probability of finding the datum $i$ given the set of parameters $p_1,p_2,\dots,p_j$. The value of $\mathcal{L}$ is usually determined over a grid that explores the region of interest in the $j$-dimension parameter space in order to identify the global maximum $\mathcal{L}(\hat{p}_1,\hat{p}_2,\dots,\hat{p}_j)$, subject to a normalization constraint that has no impact in our analysis. In practice, $\ell_i(p_1,p_2,\dots,p_j)$ is the value of the model defined by the set of parameters $p_1,p_2,\dots,p_j$  for the properties of datum $i$.

For the case of the MW satellites we are interested in, where the number of SDSS stars in each satellite can be very low (a few tens to a few hundreds over the studied region of $\pi$ deg$^2$), we restrict the set of models for the member star distribution on the sky, $\Sigma_s(r)$, to only exponential profiles. They are specified by one parameter fewer than the more customary King profiles, are also easier to deal with analytically than the Plummer models and usually yield reasonable fits\footnote{In fact, in most cases none of these profiles yields better results than the others as they do not account for the complexity that is being unveiled in an increasing number of MW satellites. Indeed several MW dwarf galaxies have now been shown to harbor multiple stellar populations (e.g. Sculptor, Fornax; see \citealt{mcconnachie07b} and references therein). However, they are a very useful and convenient way to parameterize and compare the structure of these galaxies.}. We also account for the background level of stars, $\Sigma_b$, assumed to be constant over the studied region of the sky. We therefore consider the model family $\Sigma(r|p_1,p_2,\dots,p_j)$ of stellar density in the SDSS, defined by the set of parameters $p_1,p_2,\dots,p_j$, that can be expressed as:

\begin{eqnarray}
\label{leq}
\ell_i(p_1,p_2,\dots,p_j) & = & \Sigma(r_i|p_1,p_2,\dots,p_j)\nonumber\\
 & = & \Sigma_s(r_i) + \Sigma_b\nonumber\\
 & = & S_0 \exp(-\frac{r_i}{r_e}) + \Sigma_b\nonumber\\
\end{eqnarray}

\noindent with $S_0$ the central stellar density, $r_e$ the exponential scale radius of the profile and $r_i$ the elliptical radius coordinate of datum $i$, as we consider flattened models. We are now left with the task of relating these three parameters ($S_0$, $r_i$, $r_e$) to the five parameters that need to be constrained, namely: the centroid of the satellite defined by the two celestial coordinates $\alpha_0$ and $\delta_0$ or $X_0$ and $Y_0$, the spatial offset of the centroid compared to literature values; the position angle of the satellite from north to east $\theta$; its ellipticity\footnote{Throughout this paper, the ellipticity is defined as $\epsilon=1-b/a$ with $b$ the scale-length of the system along its minor axis and $a$ that along its major axis.} $\epsilon$; its size defined by the half-light radius\footnote{The radius we determine in this manner is in fact the half-density radius but assuming there is no mass segregation in these systems, it also corresponds to the half-light radius that is commonly used in similar studies.} $r_h$. The background level $\Sigma_b$ is the sixth parameter constrained with the ML fit but its influence on $\Sigma(r)$ is obvious in equation~(\ref{leq}).

A simple integration of $\Sigma_s(r)$ out to the half-light radius easily gives the first relation $r_h=1.68r_e$. Moreover, star $i$ taken from the sample, with celestial coordinates $(\alpha_i,\delta_i)$, has a spatial position $(X_i,Y_i)$ related to the centroid by\footnote{It should be noted that this expression is only valid for stars relatively close to $\alpha_0$ and $\delta_0$, as is the case in this study. Large deviations from the centroid would require properly accounting for the projection of stars on the tangential plane of the sky.}:

\begin{equation}
\left\{   \begin{array}{lcl}
X_i-X_0 & = & (\alpha_i-\alpha_0) \cos(\delta_0)\\
Y_i-Y_0 & = & \delta_i-\delta_0\\
 \end{array} \right.
\end{equation}

\noindent while the elliptical radius is related to the spatial position by:

\begin{equation}
\label{rieq}
r_i=\bigg(  \Big(\frac{1}{1-\epsilon} (X_i\cos\theta - Y_i\sin\theta)\Big)^2 + \Big(X_i\sin\theta +Y_i\cos\theta\Big)^2  \bigg)^{1/2}.
\end{equation}

\noindent With this expression, the major axis of the system is aligned with the position angle axis.

Finally, $S_0$ is constrained by the number of stars in the SDSS sample $N_{\mathrm{tot}}$ as this number must also correspond to the integration of the stellar density $\Sigma$ over the region (of $1\deg$) around the satellite, that is, with the integration performed on the circular radius $r'$ from the centroid:

\begin{eqnarray}
N_{\mathrm{tot}} & = & \int_0^{2\pi}\int_0^{60'}\Sigma(r')dr' r' d\phi \nonumber\\
 & = & \int_0^{2\pi}\int_0^{60'}\Sigma_s(r')dr' r' d\phi +3600\pi\Sigma_b \nonumber\\
 & = & N_* + 3600\pi\Sigma_b \nonumber.\\
\end{eqnarray}

\noindent with $N_*$ the number of stars in the sample that belong to the satellite and $\Sigma$, $\Sigma_s$ and $\Sigma_b$ expressed in units of stars/arcmin$^2$. Given that, for reasonable values of $r_h$, $\Sigma_s$ is negligible at $r'=60'$, one can integrate the first term to $r'=+\infty$ instead. From there a change of variables from the circular radius $r'$ to the elliptical radius $r$ used to express $\Sigma_s$ in equations~(\ref{leq}) and~(\ref{rieq}) is required to simplify the integration. Its Jacobian is $(1-\epsilon)$, yielding:

\begin{eqnarray}
N_*  & = & \int_0^{2\pi}\int_0^{+\infty}\Sigma_s(r')dr' r' d\phi\nonumber\\
 & = & (1-\epsilon) \int_0^{2\pi}\int_0^{+\infty}\Sigma_s(r)dr r d\phi\nonumber\\
 & = & (1-\epsilon) 2\pi S_0 r_e^2
\end{eqnarray}

\noindent from which we eventually derive:

\begin{equation}
S_0=\frac{N_*}{2\pi r_e^2 (1-\epsilon)}.
\end{equation}

\noindent By substituting $r_e$, $r_i$ and $S_0$ in equations~(\ref{Leq}) and~(\ref{leq}) by their expressions, $\mathcal{L}$ is now directly expressed as a function of $(\alpha_0,\delta_0,\theta,\epsilon,r_h,\Sigma_b)$.

The set of best parameters $(\hat{\alpha_0},\hat{\delta_0},\hat{\theta},\hat{\epsilon},\hat{r_h},\hat{\Sigma_b})$ is determined through an iteratively refined grid that probes the region around the global maximum of $\mathcal{L}$. The final grid has grid point sizes of $(0.1',0.1',1.0\deg,0.01,0.1',0.0001\,\textrm{stars}/\textrm{arcmin}^2)$.

Finally, assuming the uncertainties on the various parameters are well behaved near the global maximum (i.e. Gaussian or close to Gaussian), it can be shown that $2\ln\big(\mathcal{L}(p_i)\big)$ behaves as a $\chi^2$ distribution (e.g. \S~10.3 of \citealt{lupton93}) and the $k\sigma$ confidence intervals\footnote{The confidence interval is here defined in the usual sense of a normal distribution as the regions that contain the central 68.3, 95.4, 99.73\% of the probability distribution function in the case of $1,2,3\sigma$ respectively.} of the parameters are bounded by the values that correspond to the function $2\ln\big(\mathcal{L}(p_i)\big)$ dropping by $k^2$. In practice, this translates to relative likelihood values of, respectively, 0.61, 0.136 and 0.011 for 1, 2 and $3\sigma$.

\subsection{Results}
We then proceed to apply the algorithm to consistently derive the structural parameters of the 12 Milky Way satellites recently discovered in the SDSS [BooI, Bo\"otes~II (BooII), CVnI, Canes Venatici~II (CVnII), Com, Her, LeoIV, LeoT, Segue~1 (Seg1), Ursa Major~I (UMaI), UMaII and Wil1], the 5 candidates proposed by \citet{liu08} and Objects X, Y and Z proposed by \citet{koposov08}. To test the quality of its results, we also apply the algorithm to the well studied Dra dwarf galaxy.

To do so, we extract stellar objects from the SDSS Data Release~6 \citep {adelmanmccarthy08} in a region of radius $1\deg$ around the center of the satellites. We then apply a selection box loosely constructed by eye around their CMD features ((sub-)giant branch, horizontal branch, main sequence turn-off and main sequence in the closest cases) so as to exclude clear Milky Way foreground or background contaminants. Additional magnitude cuts at $r<22.0$ and $g<22.5$ ensure that the photometry is of good quality. In the case of the very faint objects Wil1 as well as the \citet{liu08} and \citet{koposov08} candidates, we push these limits to $r<22.5$ and $g<23.0$ in order to increase the number of stars. For each satellite, the resulting stellar spatial distribution is then used to determine the best stellar density exponential model through the ML algorithm described above.

\begin{figure*}
\begin{center}
\includegraphics[width=0.8\hsize]{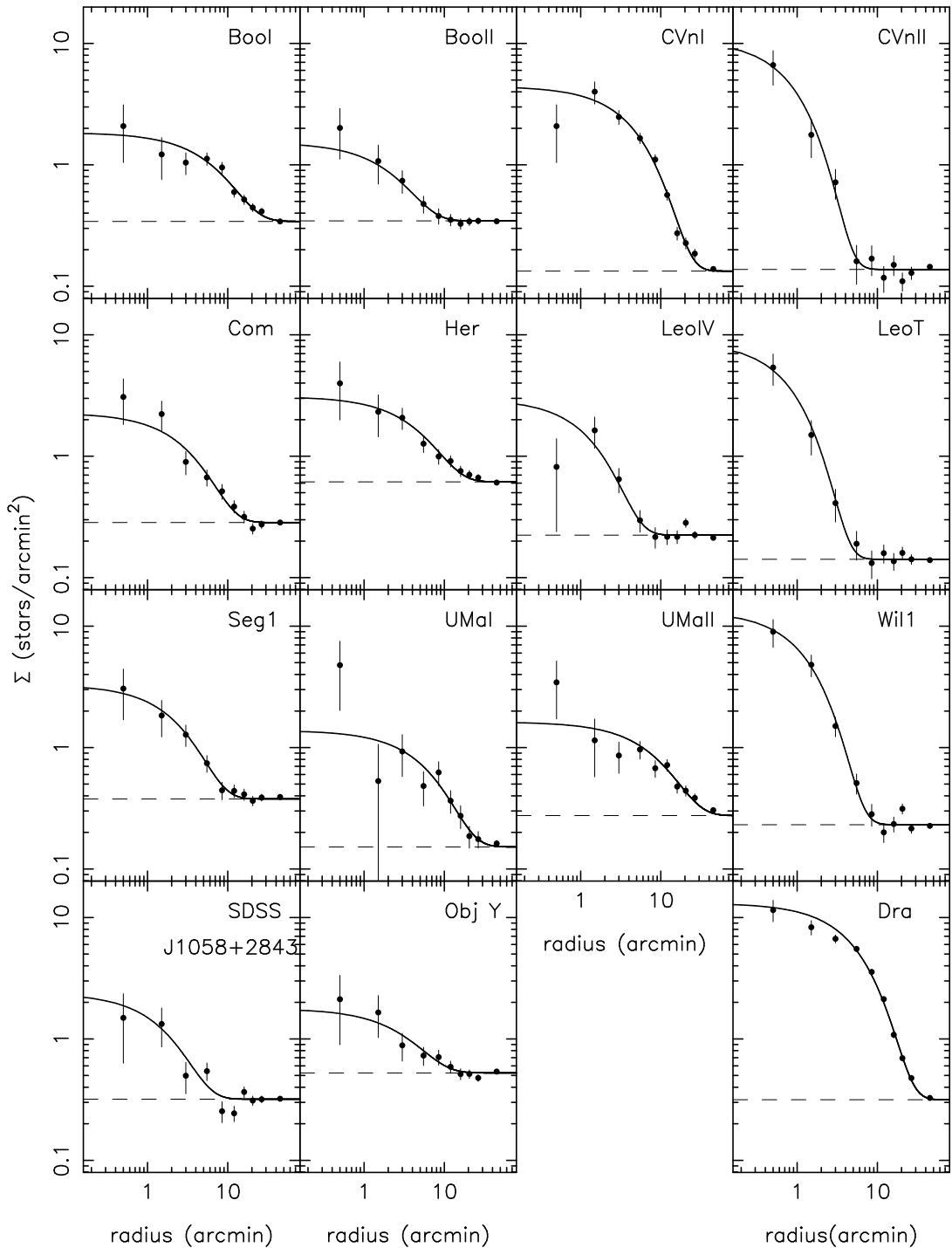}
\caption{\label{profiles}Stellar profiles of the dwarf galaxies. In all panels, the stellar density measured in fixed elliptical annuli from the SDSS data is shown as dots, using the best ellipticity, position angle and centroid found by the ML algorithm (the uncertainties are derived assuming Poisson statistics). The full lines represent the best models constructed from the parameters of Table~1. They are \emph{not} fits to the data points but show very good agreement with them, even in poorly sampled satellites such as e.g. BooII, LeoT or Seg1 that all contain less than 100\,stars.}
\end{center}
\end{figure*}

The algorithm converges, i.e. finds a unique likelihood maximum, in all cases except for 4 of the \citet{liu08} candidates and 2 of the \citet{koposov08} objects, despite the deeper magnitude limits (see below for more details). The results of the fits are summarized in Table~1 where the best values of the six parameters are listed ($\alpha_0$, $\delta_0$, $\theta$, $\epsilon$, $r_h$ and $N_*$, the integrated number of stars in the galaxy down to the assumed SDSS depth). The resulting profiles are built in Figure~\ref{profiles}, and compared to the binned data assuming the best elliptical model that is found. In all cases, the two-dimensional relative likelihood distributions are shown in the Appendix with the contours representing drops of 50, 90 and 99\% in likelihood compared to the best model. Most parameter combinations exhibit no significant covariance, except for $r_h$, $\epsilon$ and $\Sigma_b$. As expected, satellites with more stars yield better constrained models (e.g. BooI, CVnI, Dra) but even sparsely populated systems show well behaved likelihood distributions. The exception is the position angle $\theta$ in some cases, but this is expected as for (quasi-)round systems, this parameter is ill-defined (e.g. BooII, LeoIV).

\begin{table*}
\caption{Derived properties of the satellites}
{\scriptsize
\begin{tabular}{l|cccccc}
\hline\hline
 & BooI & BooII & CVnI & CVnII & Com\\
$\alpha_0$ (J2000) & 
$+14^\mathrm{h}00^\mathrm{m}03.6^\mathrm{s}\pm1.5^\mathrm{s}$ & 
$+13^\mathrm{h}58^\mathrm{m}08.0^\mathrm{s}\pm2.0^\mathrm{s}$ & 
$+13^\mathrm{h}28^\mathrm{m}03.9^\mathrm{s}\pm1.3^\mathrm{s}$ & 
$+12^\mathrm{h}57^\mathrm{m}10.0^\mathrm{s}\pm0.4^\mathrm{s}$ & 
$+12^\mathrm{h}26^\mathrm{m}57.8^\mathrm{s}\pm2.5^\mathrm{s}$ \\

$\delta_0$ (J2000) & 
$14\deg30'42''\pm43''$ & 
$12\deg50'54''\pm34''$ & 
$33\deg33'33''\pm13''$ & 
$34\deg19'33''\pm13''$ & 
$23\deg55'09''\pm19''$ \\
$\theta$ (deg.) & $+14\pm6$ & $-35^{+48}_{-55}$ & $+70^{+3}_{-4}$ & $-3^{+7}_{-9}$ & 
$-65^{+10}_{-8}$\\
$\epsilon$ & $0.39\pm0.06$ & $0.21\pm0.21$ & $0.39\pm0.03$ & $0.52^{+0.10}_{-0.11}$ &
$0.38^{+0.11}_{-0.14}$ \\
$r_h$ (arcmin) & $12.6^{+1.0}_{-0.9}$ & $4.2^{+1.1}_{-1.4}$ & $8.9\pm0.4$ & $1.6^{+0.3}_{-0.2}$ &
$6.0\pm0.6$\\
$N_*$ & $324^{+28}_{-23}$ & $37^{+11}_{-7}$ & $463\pm18$ & $29^{+5}_{-4}$ &
$99\pm13$\\
\hline
$E(B-V)$ & 0.017 & 0.031 & 0.013 & 0.010 & 0.018 \\
$D (\kpc)$ & $66\pm3$$^\mathrm{(a)}$ & $42\pm8$$^\mathrm{(b)}$ & $218\pm10$$^\mathrm{(c)}$ & $160^{+4}_{-5}$$^\mathrm{(e)}$ & $44\pm4$$^\mathrm{(d)}$\\
$r_h$ (\pc) & $242^{+22}_{-20}$ & $51\pm17$ & $564\pm36$ & $74^{+14}_{-10}$ & $77\pm10$ \\
\hline
$M_V$ & $-6.3\pm0.2$ & $-2.7\pm0.9$ & $-8.6^{+0.2}_{-0.1}$ & $-4.9\pm0.5$ & $-4.1\pm0.5$ \\
$\sigma_\mathrm{scatter}$ (mag.) & $\pm0.2$ & $\pm0.8$ & $\pm0.1$ & $\pm0.5$ & $_{-0.5}^{+0.4}$ \\
$\mu_{0,V}$ (mag/arcsec$^2$)& $27.5\pm0.3$ & $28.1\pm1.6$ & $27.1\pm0.2$ & $26.1^{+0.7}_{-0.6}$ & $27.3^{+0.7}_{-0.6}$\\
$L_V (\lsun)$ & $3.0\pm0.6\times10^4$ & $1.0\pm0.8\times10^3$ & $2.3\pm0.3\times10^5$ & $7.9^{+3.4}_{-3.7}\times10^3$ & $3.7^{+1.8}_{-1.6}\times10^3$ \\
$M_{*,\mathrm{Kroupa}} (\msun)$ & $3.4\pm0.3\times10^4$ & $1.4^{+0.7}_{-0.5}\times10^3$ & $3.0\pm0.2\times10^5$ & $8.0^{+2.0}_{-1.7}\times10^3$ & $4.8\pm0.9\times10^3$\\
$M_{*,\mathrm{Salpeter}} (\msun)$ & $6.7\pm0.6\times10^4$ & $2.8^{+1.3}_{-1.0}\times10^3$ & $5.8\pm0.4\times10^5$ & $1.6^{+0.4}_{-0.3}\times10^4$ & $9.2\pm1.7\times10^3$\\

\hline\hline
 & Her & LeoIV & LeoT & Seg1 & UMaI\\
$\alpha_0$ (J2000) & 
$16^\mathrm{h}31^\mathrm{m}05.2^\mathrm{s}\pm2.5^\mathrm{s}$ & 
$11^\mathrm{h}32^\mathrm{m}57.4^\mathrm{s}\pm2.3^\mathrm{s}$ & 
$09^\mathrm{h}34^\mathrm{m}53.6^\mathrm{s}\pm0.6^\mathrm{s}$ & 
$10^\mathrm{h}07^\mathrm{m}03.2^\mathrm{s}\pm1.7^\mathrm{s}$ &
$10^\mathrm{h}34^\mathrm{m}48.8^\mathrm{s}\pm2.8^\mathrm{s}$ \\

$\delta_0$ (J2000) & 
$+12\deg47'18''\pm17''$ & 
$-00\deg31'00''\pm26''$ & 
$+17\deg02'41''\pm9''$ & 
$+16\deg04'25''\pm15''$ &
$+51\deg56'06''\pm19''$\\
$\theta$ (deg.) & $-78\pm4$ & $-18^{+90}_{-90}$ & $-15^{+22}_{-16}$ &  $+85\pm8$ & $+71^{+2}_{-3}$\\
$\epsilon$ & $0.68^{+0.06}_{-0.08}$ & $0.22^{+0.18}_{-0.22}$ & $0.29^{+0.12}_{-0.14}$  & $0.48^{+0.10}_{-0.13}$ & $0.80\pm0.04$ \\
$r_h$ (arcmin) & $8.6^{+1.8}_{-1.1}$ & $2.5^{+0.5}_{-0.7}$ & $1.5\pm0.3$ & $4.4^{+1.2}_{-0.6}$ & $11.3^{+1.7}_{-1.3}$\\
$N_*$ & $131\pm18$ & $30\pm7$ & $31\pm5$ & $65\pm9$ & $70^{+12}_{-9}$\\
\hline
$E(B-V)$ & 0.059 & 0.024 & -- & 0.034 & 0.019 \\
$D (\kpc)$ & $132\pm12$$^\mathrm{(f)}$ & $160^{+15}_{-14}$$^\mathrm{(d)}$ & $407\pm38$$^\mathrm{(g)}$ & $23\pm2$$^\mathrm{(d)}$ & $96.8\pm4$$^\mathrm{(h)}$\\
$r_h$ (\pc) & $330^{+75}_{-52}$ & $116^{+26}_{-34}$ & $178\pm39$ & $29^{+8}_{-5}$ & $318^{+50}_{-39}$\\
\hline
$M_V$ & $-6.6\pm0.3$ & $-5.0^{+0.6}_{-0.5}$ & --\tablenotemark{(\dag)} & $-1.5^{+0.6}_{-0.8}$ & $-5.5\pm0.3$\\
$\sigma_\mathrm{scatter}$ (mag.) & $_{-0.3}^{+0.2}$ & $_{-0.5}^{+0.6}$ & -- & $_{-0.7}^{+0.6}$ & $\pm0.3$\\
$\mu_{0,V}$ (mag/arcsec$^2$) & $27.2^{+0.6}_{-0.5}$ & $27.5^{+1.3}_{-1.2}$ & -- &  $27.6^{+1.0}_{-0.7}$ & $27.7^{+0.5}_{-0.4}$\\
$L_V (\lsun)$ & $3.6\pm1.1\times10^4$ & $8.7^{+4.4}_{-4.7}\times10^3$ & -- & $335^{+235}_{-185}$ & $1.4\pm0.4\times10^4$\\
$M_{*,\mathrm{Kroupa}} (\msun)$ & $3.7\pm0.6\times10^4$ & $8.5^{+3.0}_{-2.1}\times10^3$ & -- & $600^{+115}_{-105}$ & $1.9\pm0.3\times10^4$\\
$M_{*,\mathrm{Salpeter}} (\msun)$ & $7.2^{+1.2}_{-1.1}\times10^4$ & $1.6^{+0.6}_{-0.4}\times10^4$ & -- & $1.3\pm0.2\times10^3$ & $3.7^{+0.6}_{-0.5}\times10^4$\\

\hline\hline
  & UMaII & Wil1 & SDSSJ1058+2843 & Object~Y & Dra\\
$\alpha_0$ (J2000) & 
$08^\mathrm{h}51^\mathrm{m}29.6^\mathrm{s}\pm3.3^\mathrm{s}$ & 
$10^\mathrm{h}49^\mathrm{m}21.9^\mathrm{s}\pm0.8^\mathrm{s}$ & 
$10^\mathrm{h}58^\mathrm{m}05.2^\mathrm{s}\pm1.8^\mathrm{s}$  &
$11^\mathrm{h}56^\mathrm{m}30.0^\mathrm{s}\pm5.4^\mathrm{s}$  &
$17^\mathrm{h}20^\mathrm{m}14.4^\mathrm{s}\pm0.6^\mathrm{s}$ \\

$\delta_0$ (J2000) &  
$+63\deg08'18''\pm26''$ & 
$+51\deg03'10''\pm11''$ & 
$+28\deg43'39''\pm35''$& 
$+21\deg03'06''\pm30''$& 
$+57\deg54'54''\pm8''$ \\
$\theta$ (deg.) & $-82^{+4}_{-3}$ & $+77\pm5$ & $-47^{+20}_{-24}$ & $-82^{+12}_{-22}$ & $+89\pm2$\\
$\epsilon$ & $0.63^{+0.03}_{-0.05}$ & $0.47^{+0.07}_{-0.08}$ & $0.38^{+0.17}_{-0.23}$ & $0.55^{+0.20}_{-0.16}$ & $0.31\pm0.02$\\
$r_h$ (arcmin)  & $16.0\pm1.0$ & $2.3^{+0.2}_{-0.4}$  & $3.1^{+0.7}_{-0.4}$ & $6.1^{+1.8}_{-1.2}$ & $10.0^{+0.3}_{-0.2}$\\
$N_*$ & $285^{+25}_{-23}$ & $82^{+9}_{-7}$ & $26\pm7$ & $47\pm14$ & $1930\pm34$\\
\hline
$E(B-V)$ & 0.097 & 0.013 & 0.020 & -- & 0.026 \\
$D (\kpc)$ & $30\pm5$$^\mathrm{(i)}$ & $38\pm7$$^\mathrm{(j)}$ & $24^{+2.7}_{-2.4}$$^\mathrm{(k)}$ & --\tablenotemark{(\ddag)} & $76\pm5$$^\mathrm{(l)}$\\
$r_h$ (\pc) & $140\pm25$ & $25^{+5}_{-6}$ & $22^{+5}_{-4}$ & -- &$221\pm16$\\
\hline
$M_V$ & $-4.2\pm0.5$ &  $-2.7\pm0.7$ & $-0.2^{+1.1}_{-1.0}$ & -- & $-8.75\pm0.15$\\
$\sigma_\mathrm{scatter}$ (mag.)& $\pm0.3$ &  $\pm0.6$  & $_{-1.0}^{+1.1}$ & -- & $\pm0.05$\\
$\mu_{0,V}$ (mag/arcsec$^2$) & $27.9\pm0.6$ & $26.1\pm0.9$ & $28.2^{+1.4}_{-1.3}$ & -- & $25.0\pm0.2$ \\
$L_V (\lsun)$ & $4.0^{+1.8}_{-1.9}\times10^3$ & $1000^{+660}_{-670}$  & $100^{+95}_{-100}$ & -- & $2.7\pm0.4\times10^5$\\
$M_{*,\mathrm{Kroupa}} (\msun)$ & $5.4^{+0.6}_{-0.5}\times10^3$ & $1.5\pm0.3\times10^3$ & $210^{+75}_{-60}$ & -- & $3.2\pm0.1\times10^5$\\
$M_{*,\mathrm{Salpeter}} (\msun)$ & $1.2\pm0.1\times10^4$ & $3.2\pm0.6\times10^3$  & $400^{+145}_{-120}$ & -- & $6.2\pm0.1\times10^5$\\
\end{tabular}
}
\tablecomments{
All structural parameters ($\alpha_0$, $\delta_0$, $\theta$, $\epsilon$, $r_h$ and $N_*$) are determined through the ML fit. The extinction values are derived from the \citet{schlegel98} maps and the distances are taken from the following papers: (a) \citet{dallora06}, (b) \citet{walsh08}, (c) \citet{martin08a}, (d) \citet{belokurov07a}, (e) \citet{greco08}, (f) \citet{coleman07}, (g) \citet{dejong08b}, (h) \citet{okamoto08}, (i) \citet{zucker06b}, (j) \citet{willman05a}, (k) \citet{liu08}, (l) \citet{bonanos04}. The magnitudes of the systems, $M_V$, their scatter induced by `CMD shot-noise', $\sigma_\mathrm{scatter}$, their central surface brightness, $\mu_{0,V}$, and their stellar masses, $M_{*,\mathrm{Kroupa}}$ and $M_{*,\mathrm{Salpeter}}$, are all determined as explained in \S~3.\\
(\dag) We refrain from determining the magnitude of LeoT given the presence of multiple stellar populations that render its determination very uncertain (see \S~3 for more details).\\
(\ddag) In the absence of any distance estimate for Object~Y we cannot derive parameters other than those directly estimated through the ML fit.}
\end{table*}

Below follows an object by object discussion of the derived parameters.

\emph{Dra} -- Applying our ML fitting to the Draco galaxy provides a good opportunity to verify that the algorithm works properly on a galaxy that has been well studied. Our structural parameters are very similar to previous determinations by \citet[][$\theta=88\pm3\deg$ and $\epsilon=0.29\pm0.02$]{odenkirchen01b}, also from SDSS data. The half-light radius we measure is in excellent agreement with that obtained from deeper data by \citet[][$r_h=10.09\pm0.02'$, although this value is derived from their best Plummer model]{segall07}. This comparison gives confidence in the results of the ML algorithm.

\emph{BooI} -- All the parameters derived for this galaxy are in agreement with previous measurements from \citet[][$\theta=10\pm10\deg$, $\epsilon\simeq0.33$ and $r_h=12.6\pm0.7'$]{belokurov06c}. We derive a slightly higher ellipticity but, given the uncertainties, our value is comparable to $\epsilon\simeq0.33$ (but see below for a detailed comparison).

\emph{BooII} -- No determination of structural parameters exists for this puny object except for a half-light radius of $4.0\pm1.9'$ in the discovery paper \citep[]{walsh07} and $3.6\pm1.2'$ from deeper follow-up observations \citep{walsh08}. The ML fit gives the picture of a rather spherical system.

\emph{CVnI} -- As for BooI, this galaxy is well populated in the SDSS, and the ML values are all within $1\sigma$ of those of the discovery paper \citep[][$\theta=73\pm3\deg$, $\epsilon\simeq0.38$ and $r_h=8.4\pm0.5'$]{zucker06a}.

\emph{CVnII} -- This is a faint and distant object that does not have many stars within the SDSS, which probably explains why our parameters deviate from those of \citet[][$\epsilon\simeq0.1$ and $r_h\simeq0.9'$]{sakamoto06} or \citet[][$\theta\simeq0\deg$, $\epsilon\simeq0.3$ and $r_h\simeq3.3'$]{belokurov07a}. However, since none of these authors quote any uncertainties, it is difficult to ascertain if these differences could be due to poorly constrained values. The relative likelihood contours for this satellite (Fig.~\ref{ML_plots}, $4^\mathrm{th}$ set of panels) nevertheless exhibit gentle variations and we have no reason to expect the ML algorithm to have failed and be stuck on a local maximum of the parameter space. Our best model also yields a stellar density profile that is very close to the data (Fig.~1). We find that CVnII is a rather elliptical system with an axis ratio of $\sim2:1$.

\emph{Com} -- With this reasonably populated object, we are once again in agreement with \citet[][$\theta\simeq-60\deg$, $\epsilon\simeq0.5$ and $r_h\simeq5.9'$]{belokurov07a}. We may determine the system to be slightly rounder, but $\epsilon\simeq0.5$ is just within our uncertainties.

\emph{Her} -- This galaxy is one of the recently discovered galaxies that deeper follow-up data from the Large Binocular Telescope (LBT) revealed to be more elliptical \citep[$\theta\simeq-73\deg$, $\epsilon\simeq0.65$]{coleman07} than from the smoothed maps of the SDSS \citep[$\theta\simeq-60\deg$, $\epsilon\simeq0.5$]{belokurov07a}. Since it could be the sign that this latter value is driven by the smoothing of the shallow data, it is particularly interesting that our ML value ($\epsilon=0.68^{+0.06}_{-0.08}$) is very close to the \citet{coleman07} value. Even from the SDSS data alone, we confirm that Her is one of the most elliptical satellites of the MW. On the other hand, the half-light radius we measure is similar to that of \citet[][$r_h\simeq8.4'$]{belokurov07a}. The one measured from the LBT dataset was erroneously listed as $4.4\pm0.3'$ where it should have been $9.4 \pm 1.4'$ (Matthew Coleman, private communication), once again very similar to the value of the ML fit.

\emph{LeoIV} -- This almost spherical object contains very few stars in the SDSS ($N_*=30$\,stars), leading to sizeable uncertainties. We are therefore in good agreement with \citet[][$\theta\simeq-5\deg$, $\epsilon\simeq0.25$ and $r_h\simeq3.4'$]{belokurov07a} although our fit converges towards a somewhat smaller system.

\emph{LeoT} -- Sitting in the outskirts of the MW at $\sim400\kpc$, this peculiar galaxy has until now managed to retain some gas in the form of \textsc{Hi} \citep{irwin07,ryan-weber08} and has undergone a significant episode of star formation over the last Gyr or so \citep{irwin07,dejong08b}. The SDSS data exhibit both young and old stars that are revealed to follow different density profiles from deeper data \citep{dejong08b} but given the low number of LeoT stars in the SDSS, it is necessary to merge both young and old stellar populations in the same sample to allow the ML algorithm to converge. The parameters listed in Table~1 are therefore the overall parameters of the dwarf. They are nevertheless reasonably close to those determined by \citet[][$\theta\simeq0\deg$, $\epsilon\simeq0.0$ and $r_h\simeq1.5'$]{irwin07} from SDSS data or \citet[][$r_h=1.0\pm0.1'$]{dejong08b} from deeper LBT data. In this latter case, the difference in the $r_h$ value could also be due to the different depths in the two datasets and mean that the shallow data are dominated by the more extended old population ($r_h=1.2\pm0.1'$ from the LBT data).

\emph{Seg1} -- Despite being one of the faintest MW satellites, the proximity of Seg1 to the Sun ensures it contains enough stars in the SDSS for a good fit. As for Her, we find the system to be more elliptical and slightly rotated when compared to the values determined from smoothed maps \citep[][$\theta\simeq60\deg$, $\epsilon\simeq0.3$]{belokurov07a}. The half-light radius is however in excellent agreement with their value ($r_h\simeq4.6'$).

\emph{UMaI} -- Deep data down to the main-sequence turn-off of this galaxy obtained by \citet{okamoto08} give structural parameters ($\theta\simeq78\deg$ and $r_h\simeq11.3'$) that are very similar to those yielded by the ML algorithm. We confirm the larger size of the system compared to the discovery paper \citep[$r_h\sim7.75'$]{willman05b} but, once again, derive a higher ellipticity than the one \citet{okamoto08} estimate from their smoothed maps ($\epsilon\simeq0.54$). Our value makes UMaI the most flattened MW satellite galaxy.

\emph{UMaII} -- This is yet another case were we derive an ellipticity that is higher than that determined from smoothed maps ($\epsilon\simeq0.5$; \citealt{zucker06b}). We are also able to refine the value of $r_h$ that is within the large range initially given by these authors.

\emph{Wil1} -- Table~1 gives the first measurements of  the ellipticity and position angle for this satellite, once more revealing a rather elliptical system although not as drastically so as UMaI, UMaII or Her. The ML best fit has a half-light radius similar to previous estimates from deeper data ($r_h\simeq1.75\pm0.5'$, \citealt{willman05a}; $r_h\simeq1.8'$, \citealt{martin07a}).

\emph{SDSSJ0814+5105, SDSSJ0821+5608 \& SDSSJ1329+2841} -- The ML fits do not yield significant results for these three MW satellite candidates proposed by \citet{liu08}, even when using magnitudes cuts 0.5 mag fainter. They correspond to overdensities of only $3.5$, $2.0$ and $2.5\sigma$ in $N_*$ respectively and for the last 2 objects the best ellipticity found by the algorithm is an unphysical $\epsilon=1.0$. Even though we cannot rule out that the ML is just not sensitive enough, none of these objects is visible on maps of the spatial distribution of SDSS stars, contrary to all the other objects (even the extremely faint Seg1, Wil1 or BooII).

\emph{SDSSJ1000+5730} -- From the inspection of the SDSS catalog around this object, it appears that this ``overdensity''  is in fact within $\sim25'$ of a series of holes in the photometric catalog that result from bad seeing. These holes seem to have driven \citet{liu08} to tag this underdensity of the background (that is an order of magnitude smaller than for their other candidates in their Table~2) as an overdensity produced by a MW satellite.

\emph{SDSSJ1058+2843} -- The last candidate of the \citet{liu08} list is the only one for which our algorithm converges, revealing a very sparse object with sizable uncertainties on its parameters. Our measure of the half-light radius is $1.5\sigma$ smaller than in the discovery paper but this is not surprising given the scantness of this system. Deeper photometry and/or spectroscopy are required to definitely confirm that this overdensity is indeed a MW satellite. But since its properties are similar to those of Seg1 and Wil1, we will still consider it as a satellite in our analysis (removing or keeping this satellite does not impact our overall conclusions).

\emph{Objects X \& Z} -- As for the three first candidates of \citet{liu08}, the algorithm does not converge on physical values for these two candidates proposed by \citet{koposov08}.

\emph{Object Y} -- This is the one object of the three \citet{koposov08} candidates for which the algorithm converges. Although the derived values are quite uncertain, the data and best profile agree reasonably well in Figure~1 and give the picture of a rather elliptical overdensity. This object is unfortunately too sparse for the \citet{dejong08a} CMD fitting technique to converge and the ensuing absence of any distance estimate prevents us from deriving the physical size of this object.

In summary, the ML algorithm behaves very well as is confirmed by the comparison between the preferred profiles and the binned data in Figure~\ref{profiles}. Although there are a few cases where some scatter is present, there is usually a very good correspondence between the radial profiles. Where available, the derived structural parameters are also in good agreement with previous measurements. It is particularly reassuring to notice that in the case of the well populated Draco galaxy, our parameters are very similar to those of \citet{odenkirchen01b}. In the few cases where deeper follow-up was obtained on the less populated faint satellites, the ML algorithm applied to the shallower SDSS data yields results that are consistent within the uncertainties. This is the case for the extremely faint BooII (comparison with \citealt{walsh08}) or the very elongated Her, for which we derive again an axis ratio of 3:1, as in \citet{coleman07} from LBT data reaching $\sim4$ magnitudes deeper.

The case of Her is also a good example to illustrate how smoothing the maps before deriving parameters through the intensity weighted second moments method appears to lead to measurements of smaller ellipticities: $\simeq0.5$ vs. $0.68^{+0.06}_{-0.08}$ for Her but also $\simeq0.54$ vs. $0.80\pm0.04$ for UMaI, $\simeq0.3$ vs. $0.52^{+0.10}_{-0.11}$ for CVnII, $\simeq0.3$ vs. $0.48^{+0.10}_{-0.13}$ for Seg1 or $\simeq0.5$ vs. $0.63^{+0.03}_{-0.05}$ for UMaII.

\begin{figure}
\begin{center}
\includegraphics[angle=270,width=\hsize]{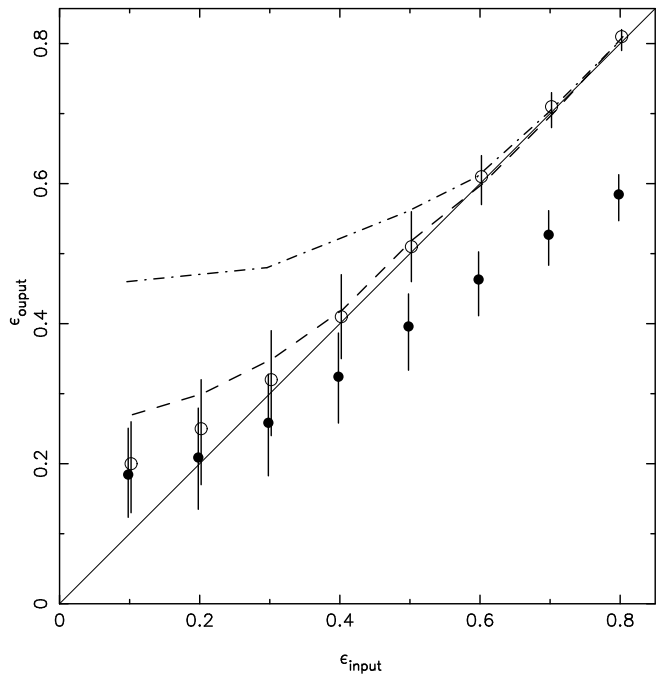}
\caption{\label{input_output_ell}Comparison of the input ellipticity of the models, $\epsilon_\mathrm{input}$, with the output ellipticity, $\epsilon_\mathrm{output}$, recovered through the intensity-weighted second moments method applied to smoothed maps (filled circles) and through the ML algorithm (hollow circles) for BooI-like systems. The points represent the median value of 500 models and the error bars on the recovered values represent the central 68.3\% of the distribution. To guide the eye, the full line represents the 1-to-1 correspondence that should be followed if the methods have no systematics: this is clearly not the case from the smoothed maps as $\epsilon_\mathrm{output}$ is systematically lower than $\epsilon_\mathrm{input}$. The dashed-dotted and dashed lines show the quality of the ML fit for lower density UMaI- and UMaII-like systems respectively. Uncertainties are similar to those of the BooI-like models but are not shown to avoid cluttering the Figure. In both cases, the recovered ellipticity falls in the region where the algorithm follows the 1-to-1 correspondence ($\epsilon_\mathrm{output}=0.80$ and $\epsilon_\mathrm{output}=0.63$ respectively).}
\end{center}
\end{figure}

To investigate this effect closer, we build 500 mock models of galaxies that are similar to BooI, one of the typical new satellites. They all have $r_h=12.6'$, $N_*=324$\,stars and an ellipticity, $\epsilon_\mathrm{input}$ that we fix. To remain as close as possible to the observed SDSS data, we also add randomly scattered background with the density found by the ML algorithm. The colors and magnitudes of these contaminants, that are chosen to fall within the selection box used to derive the properties of BooI, are randomly picked from all stars in the SDSS within a region of $5\deg\times5\deg$ around BooI, excluding the central $2\deg\times2\deg$ region. We then recover the ellipticity of the models, $\epsilon_\mathrm{output}$, using both the intensity-weighted second moments method and the ML algorithm. In the case of the former, a map of each system is built with pixels of $2'\times2'$ that is  smoothed with a round kernel of $4'$ dispersion. Only pixels that are at least $3\sigma$ over the mean background level are used (the mean background level and its r.m.s., $\sigma$, are determined from the circular annulus centered on the model and with $0.75\deg<r<0.9\deg$). For the latter, we automate the ML algorithm that searches through an iteratively refined grid. We have checked on the BooI SDSS data that the results obtained through the iteratively refined grid are very similar to those listed in Table~1 for that object. The recovered ellipticities, $\epsilon_\mathrm{output}$, are compared to $\epsilon_\mathrm{input}$ in Figure~\ref{input_output_ell} for both methods. The usefulness of the ML technique is manifest as it recovers the input ellipticity in almost all cases. Using smoothed maps, on the other hand, leads to systematically lower recovered ellipticities and this effect gets worse as $\epsilon_\mathrm{input}$ increases. Both methods have issues for the rounder systems whose recovered ellipticity is slightly higher than the input one; this effect is due to the degree of freedom afforded by $\theta$ for the (quasi-)round systems in the presence of a few outer stars. 

This example shows the limitations of smoothed maps in determining the structural parameters of faint satellites. Not only do they require some fine-tuning to have a set of pixel size, kernel size and background threshold that translate into a suitable map, but even with such a map, the ellipticity is measured rounder than it truly is. This effect becomes even more important as the number of stars in the system shrinks. The strongest deviations are indeed observed for systems that are less populated than BooI: CVnII, Her, UMaI or Seg1. Since in most cases, this $\epsilon$ is then used to construct the radial profile of the satellites, this can have a significant impact on the derived structure of the system. In contrast, the ML fit is directly applied to the data (the spatial distribution of stars), and all the parameters are derived at the same time, leading to best values and uncertainties that include the influence of the determination of the other structural parameters.

While this analysis of the behavior of the ML fit is only done in detail for BooI-like systems, it should be noted that the efficiency of the algorithm actually scales with the density of stars in the satellite and not solely with their number of stars $N_*$ within the SDSS. Since a lower limit of the density $\rho$ can be estimated by assuming a round morphology, it scales as $\rho\propto\frac{N_*}{r_h^2}$. A look at Table~1 shows that $\rho$ is higher than that of BooI in all other cases, except for UMaI, UMaII and Object~Y. Therefore, apart from these three systems, the quality of the ML fit remains at least as good as shown by the hollow circles in Figure~\ref{input_output_ell} and only very round systems can be found to be slightly more elliptical than they really are. The cases of the less dense UMaI and UMaII are also shown in the Figure as the dashed-dotted and dashed line respectively (the dashed line also applies to Object~Y as its density is similar to that of UMaII). Even though the overestimate of small ellipticities is more significant than previously, the ellipticities of these three systems are so high that they should not be hampered by the algorithm: in both cases, the recovered ellipticity falls in the region where the algorithm behaves properly and follows the 1-to-1 line.

Generally, the over-prediction of the ellipticity of very round systems could become a difficulty in the case of systems sparser than those considered here. But recently discovered satellites all lie at the detection limit of the SDSS \citep{koposov08} and most fainter (and sparser) candidates, such as the \citet{liu08} and \citet{koposov08} ones, are not dense enough to allow the algorithm to converge. Their confirmation definitely requires deeper datasets than the SDSS, leading to denser samples on which to apply the ML technique. This will consequently elevate the issue of the overestimation of small ellipticities.

\section{Luminosities and stellar masses}
The use of the total magnitude to quantify the amount of stars in these very sparsely populated systems is problematic. Their observed luminosity can indeed vary strongly through time, depending on whether a few stars have evolved along the red giant branch or most of them remain along the main sequence or main sequence turnoff. To remove the influence of this `CMD shot-noise', we propose to characterize the systems by their \emph{number} of member stars (above a luminosity threshold), $N_*$, instead of the \emph{luminosity} of these stars, and study the impact of $N_*$ on the total magnitudes and stellar masses that are likely to be observed.

For each satellite\footnote{Except for Object~Y, whose lack of a distance estimate prevents us from determining its absolute magnitude.} we have stellar population models, also derived from SDSS data \citep{dejong08a}, and the ML fit gives $N_*$ with its associated uncertainty; so we can generate numerous CMDs of each system. The luminosity of the stars in each of these CMDs then yields the `observed magnitude' of \emph{this particular realization} of the satellite. We then define the `expected absolute magnitude' of the system as the median of the distribution of observed magnitudes\footnote{As these distributions are close to being normal distributions, this is almost equivalent to taking the mean of the distributions.}. In detail, a high signal-to-noise CMD is created for each dwarf, containing on average 70\,000 stars, which includes a realistic photometric error and completeness model for SDSS data \citep{dejong08a}. These `master-CMDs' are populated assuming a standard Salpeter or Kroupa initial mass function \citep{salpeter55,kroupa93} down to the hydrogen burning limit (0.08 $\msun$) and a binary fraction of 0.5\footnote{In Draco, \citet{kleyna02} have shown that there is mild evidence of a lower binary fraction than the solar neighborhood value used here. However, assuming a fraction of 0.3 does not significantly change our derived magnitudes: in the case of UMaII which, owing to its proximity, has one of the deepest CMDs of all the satellites, the measured magnitude changes by less than 0.05.}. Since all the satellites considered here are dominated by old populations, the stellar ages are spread between 10 and 13 Gyr in all cases and  the metallicities are chosen to correspond to the best-fit values from \citet[][given in their Table~5]{dejong08a}, where available. As these metallicities are only a means to parametrize and reproduce the SDSS CMDs, we abstain from using spectroscopic metallicities. Despite being more accurate, these could be systematically offset from the isochrones and lead to artificial CMDs that are different from the observed ones. In the case of BooII, Draco and SDSSJ1058+2843, whose stellar population models were not derived by \citet{dejong08a}, we assume the same old populations and  metallicities of $\FeH=-2.0$ \citep{walsh07}, $-1.8$ \citep{aparicio01} and $-2.0$ \citep{liu08} respectively. In order to track not only the luminosities of the satellites, but also their stellar masses, the mass of each star in the `master-CMDs' is stored, together with its magnitude and color.

One thousand realizations of each satellite are generated randomly by selecting $N_*$ stars (with its associated uncertainty listed in Table~1) from the `master-CMD' to fall within the selection box previously used to select member stars. The `observed magnitude' of each realization is then determined from the luminosity of all stars above a limiting magnitude of $g=21.5$ (half a magnitude brighter than the approximate 100\% completeness limit of the SDSS), transformed to the $V$-band using color equations from \citet{jester05}. Using the distance listed in Table~1, the luminosity of stars below the magnitude limit is accounted for by using luminosity function (LF) corrections derived from the LFs of \cite{dotter07} with the metallicity and age closest to each model population.

In parallel, we also determine the stellar mass of the satellites by determining the mass down to the same magnitude limit as   above and then correcting for the `unobserved' stars present in the `master-CMDs', down to the hydrogen burning limit. This correction\footnote{In the stellar mass range probed by the SDSS photometry of these old populations, the slope of both IMFs is the same, so that this extrapolation is valid in both cases.} is done for both a Salpeter and a Kroupa IMF to yield two estimates of the stellar mass listed in Table~1 for each satellite ($M_{*,\mathrm{Salpeter}}$ and $M_{*,\mathrm{Kroupa}}$ respectively).

\begin{figure*}
\begin{center}
\includegraphics[angle=270,width=0.49\hsize]{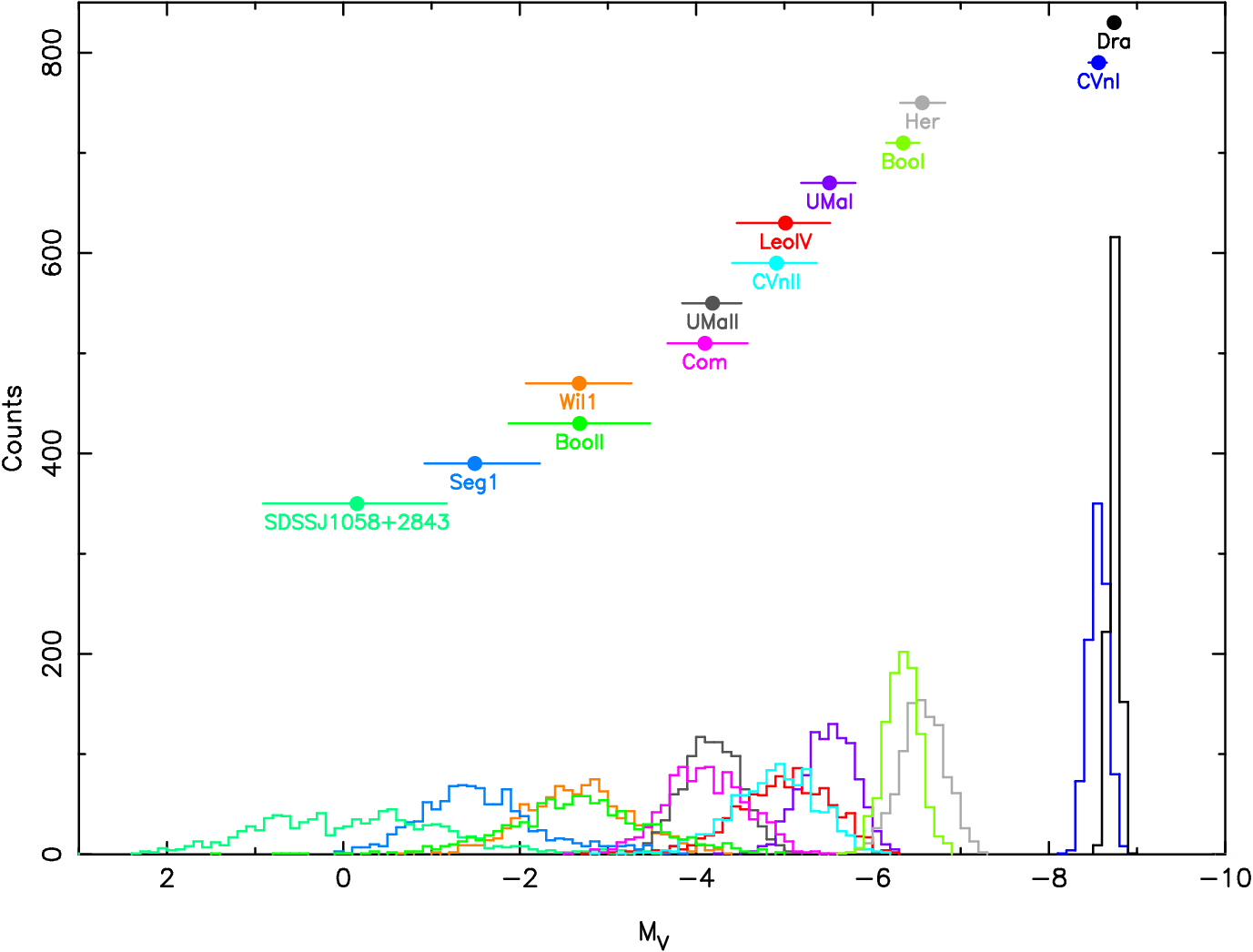}
\includegraphics[angle=270,width=0.49\hsize]{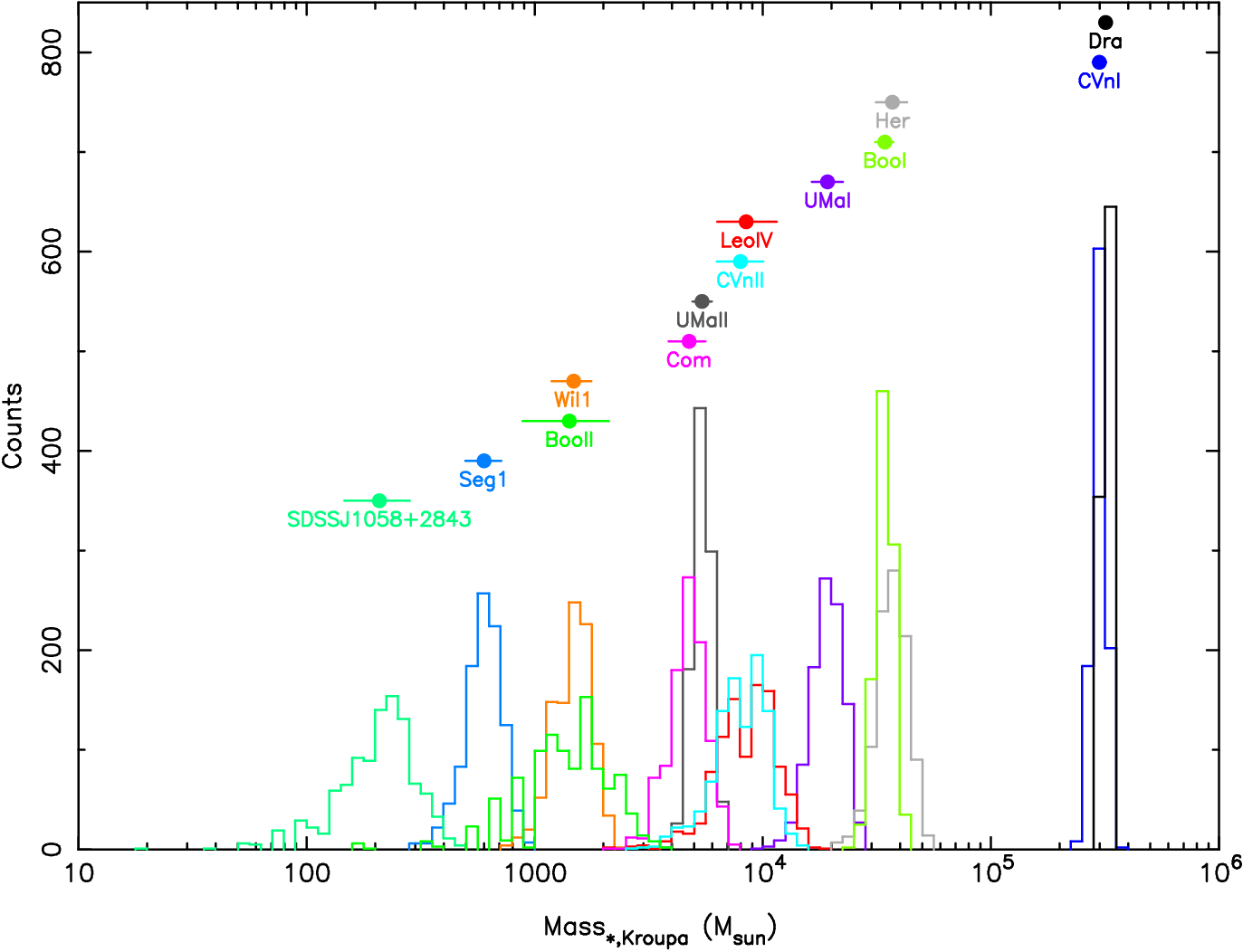}
\caption{\label{fig_Mv}Impact of the `CMD shot-noise' on the `observed magnitude' and `observed stellar mass' of the satellites. Each histogram corresponds to a different galaxy and shows the distributions of `observed magnitudes' (left) and `observed stellar mass' (right) for different realizations of their CMD. The median value of each distribution, the `expected absolute magnitude' (resp. `expected stellar mass') that best characterizes the luminosity (resp. mass) of the system, is shown as a filled dot and the horizontal error bar represents their $1\sigma_{\mathrm{scatter}}$ (the region corresponding to the central 68.3\% of the distributions). As expected, faint satellites show a much larger scatter in their magnitude since the presence/absence of a few very bright red giant branch stars will strongly impact the recovered value. The masses show a smaller scatter as they are less sensitive to the evolution of the stars.}
\end{center}
\end{figure*}

The resulting distributions of these `observed magnitudes' and `observed stellar masses' are shown in Figure~\ref{fig_Mv} along with their `expected absolute magnitude' and `expected stellar mass', taken to be the median value of the distributions. The scatter produced by the `CMD shot-noise' is readily visible from the horizontal error bars that represent the central 68.3\% of the distributions. All values and their $\pm1\sigma_{\mathrm{scatter}}$ are listed in Table~1. As could be expected, bright galaxies such as CVnI and Dra have an extremely small scatter, as their $N_*$ is very well determined by the ML fit. $N_*$ is also large enough that there is a good sampling of the CMD: whatever the random realization, the resulting CMD is very similar. On the other hand, as the satellites become fainter and fainter, their CMDs are only sparsely sampled for each realization and the resulting magnitude is strongly influenced by the presence of only a few very bright stars. The stellar masses are much less affected by these CMD sampling problems; indeed the evolution of an individual star does not change its mass as strongly as its luminosity.

For a proper comparison of our absolute magnitudes with the literature, we need to include the uncertainties that come from our assumptions on the age and metallicity of the dwarf and thence on the LF correction. This correction is, however, almost insensitive to $\pm0.5$\,dex changes in the metallicity (less than 1\% change) while a shift of $\pm2\Gyr$ yields small changes of only $\sim3\%$ in all cases. This translates into uncertainties of less than $\sigma_{\mathrm{LF}}=0.06$\,mag. Another source of uncertainty can come from the assumption that the satellites are well represented by a single population model. In fact we know in the case of CVnI that this is not the case as $\sim5\%$ of the stars are found to be young \citep{martin08a}. However, such a low proportion of young stars only minutely affects the luminosity correction (by less than 1\%) and does not impact our results. This is not the case anymore for systems with a significantly extended star formation history and we therefore refrain from determining a magnitude for LeoT.

Finally, we also account for the uncertainty coming from the distance measurements, $\sigma_{\mathrm{dist}}$ (usually of the order of 10\% in distance or 0.3\,mag; see Table~1). The final uncertainties given in Table~1 correspond to the quadratic sum of all three uncertainties: $(\sigma_{\mathrm{scatter}}^2+\sigma_{\mathrm{LF}}^2+\sigma_{\mathrm{dist}}^2)^{1/2}$. Comparison of these with values of $\sigma_\mathrm{scatter}$ readily shows that the `CMD shot-noise' is the main driver of uncertainties on the `expected absolute magnitude' of such faint systems.

In the end, almost all of our luminosity measurements are less than $1\sigma$ from literature values but with slightly smaller error bars. The Draco value is in excellent agreement with that of \citet[][$M_V=-8.8\pm0.2$]{mateo98}. We confirm the extremely low luminosity of BooII, UMaII, Wil1 and Seg1, although for this latter case, we find it is even fainter than previously measured ($-1.5^{+0.6}_{-0.8}$ vs. $-3.0\pm0.6$). At the other end of the spectrum, we find that CVnI is somewhat brighter than previously inferred, which makes it almost as bright as Draco: this galaxy should probably not be seen as a peculiar faint system but mainly owes its late discovery to its large distance. Our value for UMaI is significantly smaller than that of the discovery paper \citep[][$M_V\simeq-6.75$]{willman05b} but, as for its half-light radius, it confirms the new measurement by \citet[][$M_V\simeq-5.5$]{zucker07b}. This value also solves the issue of BooI containing more stars than UMaI while being an apparently fainter system \citep{munoz06b}: BooI \emph{is} brighter than UMaI and \emph{should} contain more stars.

From the luminosity and structural parameters of the systems, we also estimate their central surface brightness $\mu_{0,V}$, listed in Table~1. As the central surface brightness (expressed in units of $\lsun$/arcsec$^2$) is proportional to $(1-\epsilon)^{-1}$ [see equation (6)], our newly derived parameters sometimes have a strong impact on the derived $\mu_{0,V}$ that can be significantly different from previous estimates. When our morphological model is similar to previous models, we nevertheless reach very similar values (e.g. Dra with $\mu_{0,V}=25.0\pm0.2$\,mag/arcsec$^2$ vs. $\mu_{0,V}=25.3\pm0.5$\,mag/arcsec$^2$ in \citealt{mateo98} or BooI with $\mu_{0,V}=27.5\pm0.3$\,mag/arcsec$^2$ vs. $\mu_{0,V}=27.8\pm0.5$\,mag/arcsec$^2$ in \citealt{belokurov06c}). We find that there is no system fainter than $\mu_{0,V}\simeq28.0$\,mag/arcsec$^2$, a threshold that is brighter than previously claimed but in very good agreement with the detection limit of the \citet{koposov08} automatic search.

Apart from the scatter due to the uncertainty in $N_*$, the largest contribution to the uncertainty of any stellar mass determination comes from the assumed IMF. The strength of this effect is readily visible by comparing the two stellar mass estimates listed in Table~1 for the Salpeter and Kroupa IMFs.

It should be noted that even in the case of very faint satellites, the changes in the luminosities of the systems have only a small impact on the determinations of their total (dark and luminous) mass \citep{kleyna05,munoz06b,martin07a,simon07,strigari07} as these are very model-dependent and strongly dominated by other uncertainties.

\section{Are the stellar distributions `distorted'?}
\begin{figure*}
\begin{center}
\includegraphics[angle=270,width=\hsize]{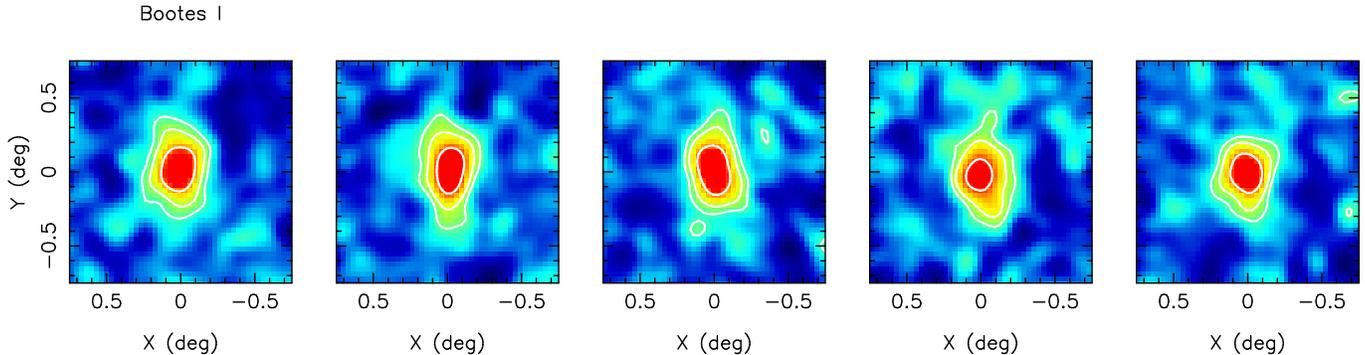}
\caption{\label{fig_maps}Smoothed maps of BooI from the SDSS data (left) and four simulations of BooI constructed assuming the best parameters of Table~1. The smoothing kernel is a two-dimensional Gaussian of dispersion $4'$. The contours correspond to detections of $3\sigma$, $5\sigma$ and $10\sigma$ over the background level, measured in an annulus between $0.75\deg$ and $0.9\deg$. Although BooI has a distorted morphology, the purely spheroidal simulated models show as much distortion.}
\end{center}
\end{figure*}

The best-fit model of a satellite can also be used to test whether the data imply deviations from this best (smooth and symmetrical) model. One of the striking features of most of the recently discovered systems is their apparent clumpiness, even when the stellar distribution is smoothed. It has been suggested on numerous occasions that their distorted isophotes could be the sign of tidal distortion or intrinsic clumpiness \citep{belokurov06c,zucker06b,willman06,belokurov07a,coleman07,martin08a}. However, the (very) low number of detected stars may leave the same impression, even if the parent distribution were smooth and symmetric. A good visual example of such a `low number statistic' behavior is given in Figure~\ref{fig_maps}, where the smoothed map of BooI (left) is compared to similar maps built from \emph{purely spheroidal} models with the structural parameters obtained from the ML algorithm $\epsilon=0.39\pm0.06$, $r_h=12\mcnd6_{-0.9}^{+1.0}$ and $N_*=324^{+28}_{-23}$ (the background contamination is accounted for in the same way as in the models presented in \S~2). It is readily apparent that, even with the smoothing that tends to drown small clumps, the realizations of purely spheroidal models appear just as distorted as the observed data. Taken at face value, most of them would in fact be suspected to contain `tidal features'. \citet{walsh08} reach similar conclusions in their analysis of BooII.

To better estimate the (non-)significance of the apparent distortions, we determine, for all satellites, the fractional r.m.s. deviation $\sigma_\mathrm{sc}/\textrm{total}$ of the data compared to the best model, accounting for the expected contribution of Poisson counting uncertainties. This is similar to the method that \citet{bell07} applied to measure the clumpiness of the Milky Way halo. Although the reader is referred to this paper for a detailed description, the gist of the method is to compare the actual deviations between the data and the model with the expected deviations from Poisson statistics. For a binned (but not smoothed) map of the considered system with $N$ pixels of interest we therefore compute:

\begin{eqnarray}
\lefteqn{\Big(\frac{\sigma_\mathrm{sc}}{\textrm{total}}\Big)^2 =} \nonumber\\
& & \Big(\frac{1}{N}\sum_{i=1}^N(D_i-M_i)^2 - \frac{1}{N}\sum_{i=1}^N(P_i-M_i)^2\Big) . \Big(\frac{1}{N}\sum_{i=1}^ND_i\Big)^{-2},
\end{eqnarray}

\noindent where $D_i$ is the number counts in the data for pixel $i$, $M_i$ the expected number of stars in this pixel from the best model determined in Table~1 and $P_i$ that of a Poisson realization of this best model. To avoid being sensitive to clumpiness in the background, induced e.g. by the clustering of misidentified background galaxies, we only consider the $N$ pixels for which $M_i$ is at least twice the number of stars expected from the background alone. In contrast to \citet{bell07}, who have a large number of pixels, the value of $N$ remains very low in the case of the maps of Milky Way satellites. To still get a good handle on the influence of Poisson noise, we rely on a Monte-Carlo analysis and generate numerous Poisson realizations of the models to study the \emph{distribution} of $(\sigma_\mathrm{sc}/\textrm{total})^2$ values for each system. Moreover, since these values can become negative in the case of an absence of distortion, we refrain from determining $\sigma_\mathrm{sc}/\textrm{total}$ right away. In this case, $(\sigma_\mathrm{sc}/\textrm{total})^2<0$ means that the scatter in the SDSS data is actually smaller than the typical scatter expected from shot-noise for the best model.

\begin{figure*}
\begin{center}
\includegraphics[angle=270,width=0.33\hsize]{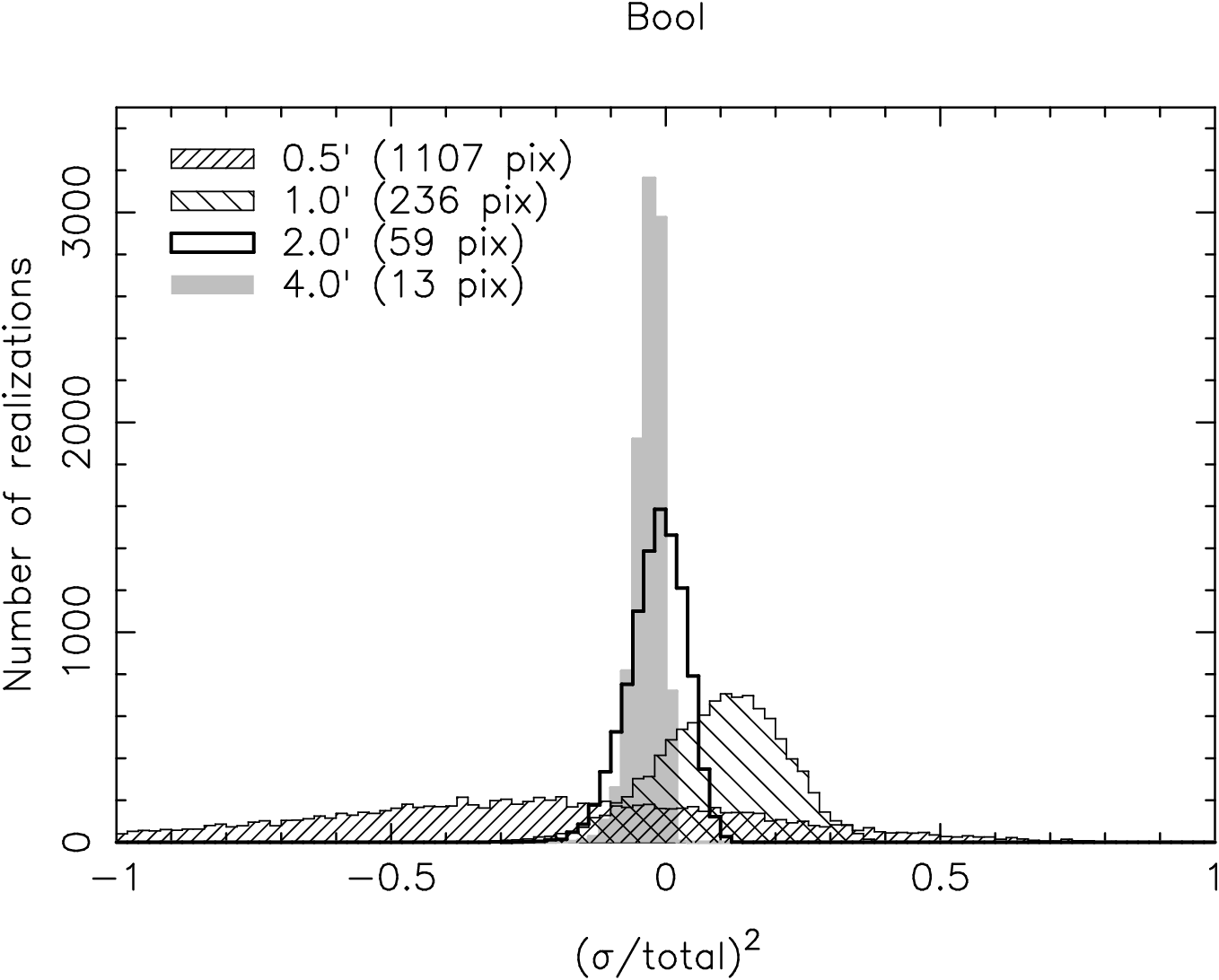}
\includegraphics[angle=270,width=0.33\hsize]{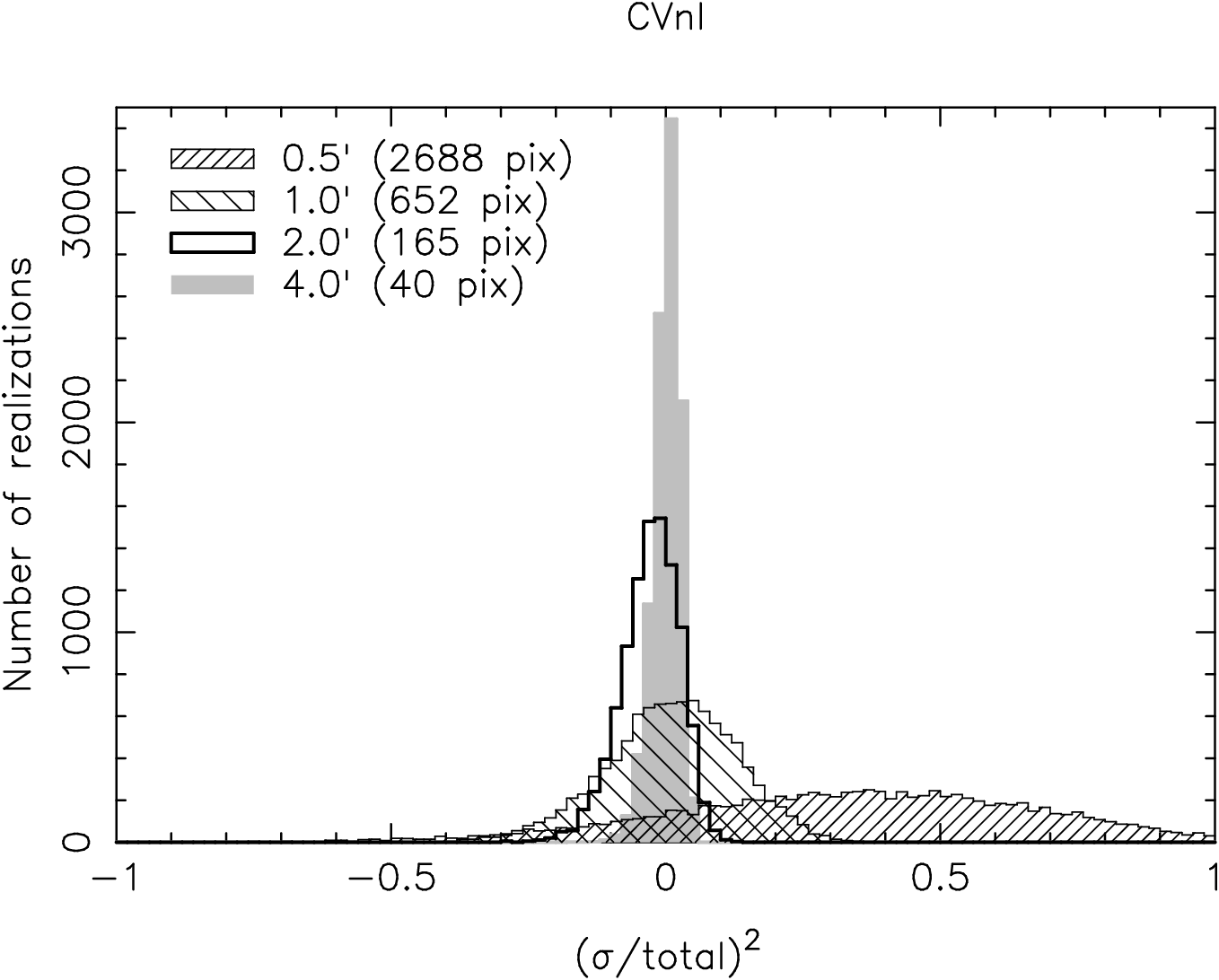}
\includegraphics[angle=270,width=0.33\hsize]{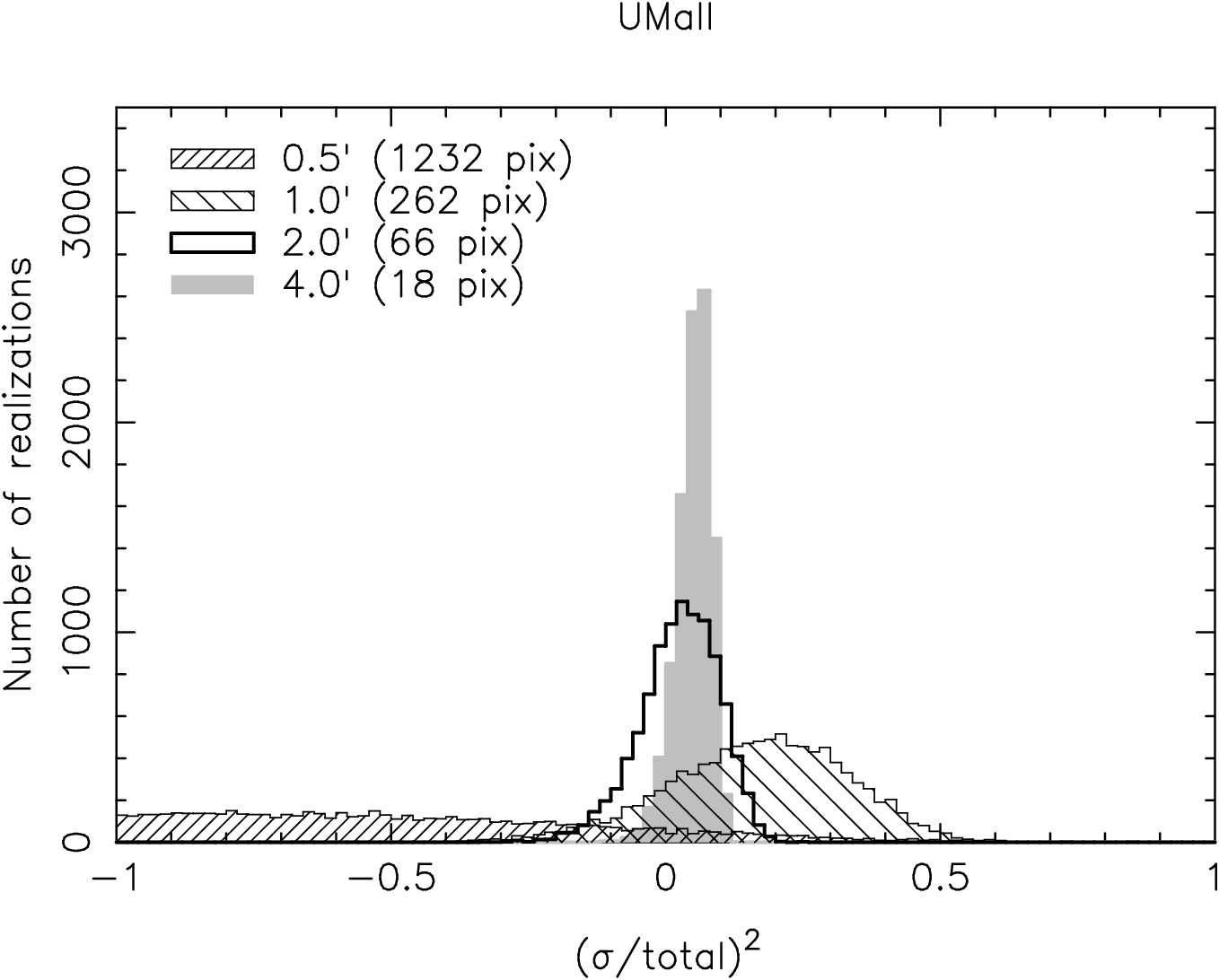}
\caption{\label{rms_fig}Distributions of $(\sigma_\mathrm{sc}/\textrm{total})^2$ for the three dwarf galaxies BooI (left), CVnI (middle) and UMaII (right). In each case, the results are shown for four grids with pixel sizes of $0.5'$ (left-dashed histogram), $1'$ (right-hashed histogram), $2'$ (white histogram) and $4'$ (grey histogram). Only the distributions obtained for CVnI with the $0.5'$ grid and for UMaII for the $1.0'$ and $4.0'$ grids are different from 0 at least at the $1\sigma$-level. The distributions obtained for BooI are typical of those obtained for the other dwarf galaxies.}
\end{center}
\end{figure*}

As noted by \citet{bell07}, this method is insensitive to the chosen pixel scale but will only be able to measure the influence of deviations from the best model that have a size at least bigger than the chosen pixel size. We therefore assume 4 different grids with pixel scales of $0.5'$, $1'$, $2'$ and $4'$ for which we determine the distribution of $(\sigma_\mathrm{sc}/\textrm{total})^2$ values\footnote{It should be noted, however, that for close objects, the grid with the smallest pixels is not very reliable as most pixels are expected to contain either 0 or 1 star. This grid is only useful for the distant ($D\gta50\kpc$) and populated objects in our sample.}. The derived distributions are shown in Figure~\ref{rms_fig} for three examples: BooI where the pixels correspond to physical sizes of 10, 19, 38 and $77\pc$, CVnI (respectively 32, 63, 127 and $254\pc$) and UMaII (respectively 4, 9, 17 and $35\pc$). The influence of the pixel grid is readily visible: the distributions derived from the small pixel grids have much more scatter, while larger pixel grids lead to almost no scatter and are closer to zero. In that case, there are not enough pixels for such sparse systems (and these examples are among the most populated of the new MW satellites).

In most cases the distributions are similar to the ones of BooI shown on the left panel and imply the absence of any significant deviation from a scatter in the stellar distribution that is only produced by shot-noise. However, CVnI and UMaII are the two cases where Poisson scatter alone may not account for the measured scatter in the distributions. The distribution from the $0.5'=32\pc$ grid of CVnI yields $(\sigma_\mathrm{sc}/\textrm{total})^2=0.35^{+0.32}_{-0.34}$, a mild $1\sigma$ detection that might be induced by deviations from the best ML model in the data. There is a slightly more significant deviation in UMaII with the $1.0'=9\pc$ and $4.0'=35\pc$ grids yielding respectively $(\sigma_\mathrm{sc}/\textrm{total})^2=0.27^{+0.14}_{-0.17}$ and  $(\sigma_\mathrm{sc}/\textrm{total})^2=0.06\pm0.03$, or a contribution of intrinsic clumpiness to the scatter of $52^{+12}_{-20}\%$ and $24\pm7\%$ (the results of the $2'$ grid, $22_{-22}^{+13}\%$, although consistent with no distortion, have large enough uncertainties that they are also consistent with a signal of the order of that found for the $1'$ and $4'$ grids).

It is reassuring that the two dwarfs that show the highest evidence of clumpiness with this method are the two cases where distortions or substructure have been argued on other grounds. Indeed, a clump of young stars was uncovered in deep LBT observations of CVnI \citep{martin08a} and, with a size of $\sim1'$, it should only be picked up by the grid with the smallest pixel size. In addition, UMaII, the other galaxy that appears to contain a deviation from its best model is the likely progenitor of the Orphan Stream and is therefore expected to be suffering from tidal disruption (\citealt{fellhauer07,martin07a} although see \citealt{sales08}). On the other hand, these two cases are only $1-2\sigma$ deviations from null-detections and, given the 60 distributions generated for the 15 studied satellites, it cannot be ruled out that they are in fact chance detections that have nothing to do with real distortions in the morphology of the dwarfs. Although tentative, the detections in CVnI and UMaII therefore need to be confirmed from deeper, more populated datasets. In all the other satellites, the scatter in the stellar density can be entirely accounted for by shot-noise in the SDSS data. Either these systems intrinsically contain no distortion or the SDSS is too shallow to allow detection of any such clumpiness.

\section{Discussion}
We have re-derived the structural parameters and magnitudes for all the newly discovered Milky Way satellites (Table~1). Not only are these determined in a homogeneous manner for all objects, but the ML algorithm produces robust results, even for extremely faint systems that only contain a few tens of stars in the SDSS. As mentioned above, our results are overall in good agreement with previous measurements although there are some systematic differences, such as a higher ellipticity found with the ML technique compared to that measured from smoothed maps. Those cases are easily understood since sparsely populated smoothed maps tend to be dominated by the (round) smoothing kernel.

\begin{figure*}
\begin{center}
\includegraphics[angle=270,width=0.8\hsize]{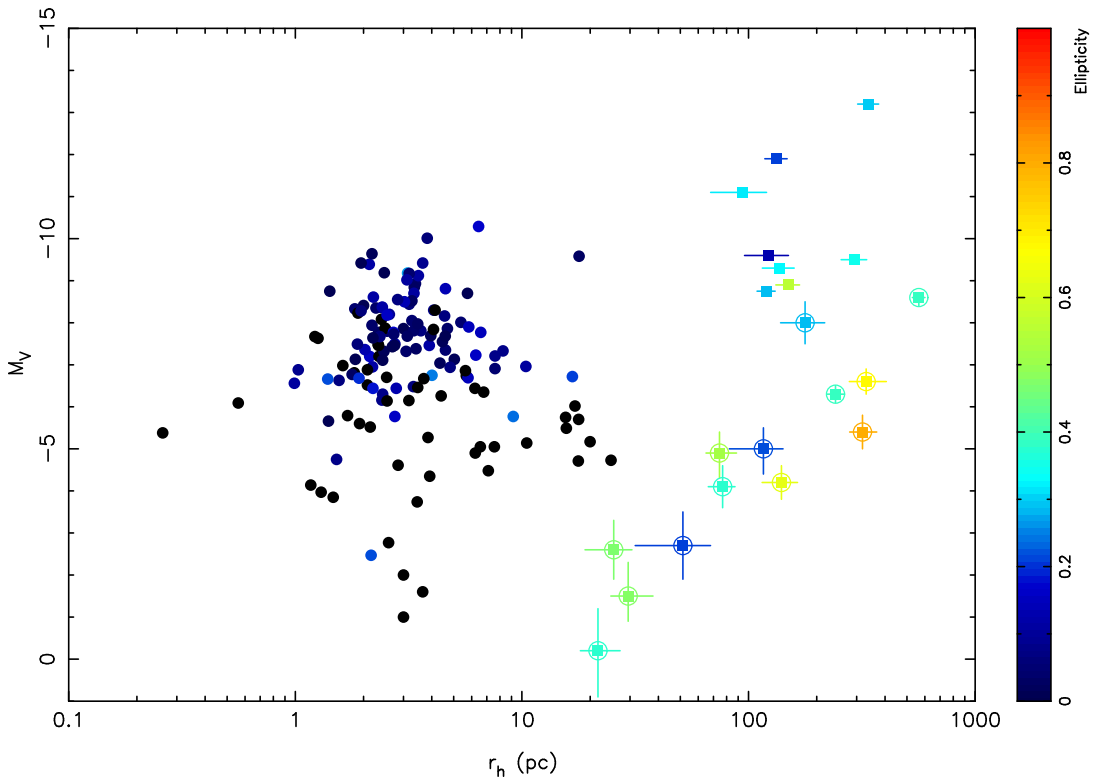}
\caption{\label{Mv_rhb}Distribution of the MW satellites in the $r_h$-$M_V$ space. Colors code their ellipticity as defined on the wedge and black is used when no ellipticity measurement is available. Globular clusters, taken from \citet{mackey05} and \citet{koposov07}, are shown as big dots and are rather small systems clustered on the left side of the plots; previously known, bright, dwarf galaxies are shown as filled squares \citep{mateo98}, while new MW satellites appear as circled squares and become fainter and smaller than the other dwarfs, seemingly bridging the size gap that appears at higher luminosity between GCs and dwarf galaxies. Error bars represent the uncertainties taken from Table~1 or the literature. The region of large and faint objects (bottom right) is incomplete due to surface brightness limits from the SDSS \citep{koposov08}.}
\end{center}
\end{figure*}

The complete dataset can be summarized on the $r_h$ vs. $M_V$ plane where dwarf galaxies and globular clusters were well separated prior to recent discoveries. Figure~\ref{Mv_rhb} shows the Milky Way satellite system with globular clusters taken from the list of \citet{mackey05} (with the addition of the two extremely dim clusters found by \citealt{koposov07}) represented as big dots, previously known dwarf spheroidal galaxies\footnote{These include Carina (Car), Dra, Fornax (For), Leo~I, Leo~II, Sculptor (Scu), Sextans (Sex) and Ursa Minor (UMi). The Sagittarius galaxy (Sgr) was removed from the sample as it is clearly interacting with the Milky Way and its properties are quite uncertain. As the only transitional galaxy of the sample, LeoT could arguably be removed from the following analyses, but that would not change our results given the mild (and uncertain) ellipticity of this dwarf.} as filled squares (the values are taken from \citealt{mateo98}) and recently discovered SDSS satellites as circled squares. All the new satellites are smoothly distributed from the region inhabited by the brighter dwarf galaxies, down to extremely small and faint systems. As previously noted by \citet{belokurov07a} and \citet{gilmore07}, the clear size difference that existed before between globular clusters, with sizes smaller than $\sim20\pc$ and dwarf galaxies $\gta100\pc$ becomes blurred at low luminosities. Many SDSS satellites are in fact smaller that $100\pc$ (BooII, CVnII, Com, SDSSJ1058+2843, Seg1 and Wil1). Therefore, it might appear more difficult to assign these systems to one population or the other. However, spectroscopic surveys conducted so far \citep{martin07a,simon07} show that they are dark matter dominated (except maybe for SDSSJ1058+2843 and Seg1 for which no spectroscopic data have yet been published).

Another obvious feature of Figure~\ref{Mv_rhb} is the higher ellipticity of dwarf galaxies compared to globular clusters. Unfortunately, most faint GCs ($M_V\gta-6.0$) have no ellipticity measurement in the \citet{mackey05} list. They also either lie away from the region covered by the SDSS or have incomplete photometry due to crowding at their center. As a result, we cannot determine their structural parameters using the ML algorithm. Still, in cases where crowding leads to incomplete data in the SDSS catalog, the clusters do appear visually round in their outskirts. Pal~5 ($M_V=-5.2$) is one of the few GCs to which we can apply the ML algorithm. This object is of particular interest as it is known to be tidally disrupting in the gravity field of the Milky Way and shows beautiful stellar tidal tails \citep{odenkirchen01a,rockosi02,grillmair06}. Even in such an extreme case when one could expect the measured ellipticity to be inflated by the tails, applying the ML algorithm yields $\epsilon=0.11\pm0.04$ and the cluster remains very round. Moreover the 5 faint GCs in the list that have an ellipticity measurement all have $\epsilon<0.25$ so there does appear to be a significant difference between the flattening of genuine GCs and that of newly discovered MW satellites.

\begin{figure}
\begin{center}
\includegraphics[angle=270,width=\hsize]{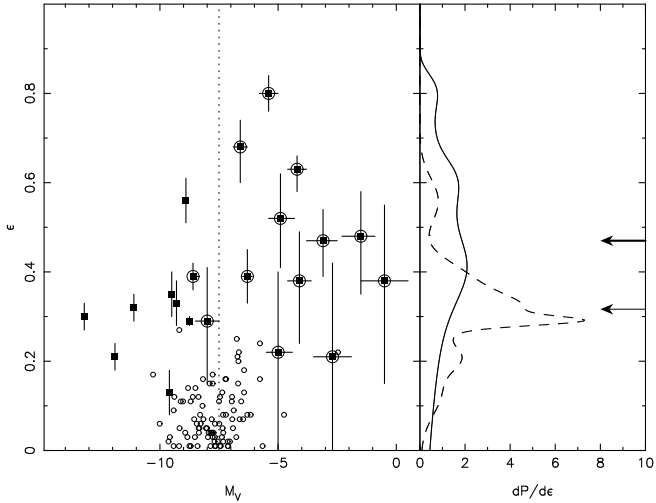}
\caption{\label{ell_fig} Evolution of the ellipticity of MW satellites as a function of their absolute magnitude (left panel). Globular clusters are this time shown as small hollow circles to avoid cluttering the plot, while the symbols for dwarf galaxies are the same as in Figure~\ref{Mv_rhb}. Error bars correspond to the uncertainties listed in Table~1 or in \citet{mateo98} for the brighter galaxies. Dividing the complete sample at $M_V=-7.5$ (dotted line) leads to the probability distribution function (PDF) of the faint subsample, represented by a full line in the right panel, that extends to higher $\epsilon$ values than that of bright galaxies (dashed line). The mean of the PDF of faint (bright) systems is represented by the thick (thin) arrow.}
\end{center}
\end{figure}

The left panel of Figure~\ref{ell_fig} shows the distribution of MW satellites in the $\epsilon-M_V$ plane. As was already reported by \citet{vandenbergh08} using the previous, rounder ellipticity estimates, GCs and dwarf galaxies populate different regions of the plot (a Kolmogorov-Smirnov (KS) test actually yields a probability of 0\% that both datasets follow the same ellipticity probability distribution). Such a drastic difference is probably related to the formation mechanism of the objects in the two samples and/or to the presence of a dark matter halo around dwarf galaxies whereas GCs are not believed to contain any dark matter. As for the dwarf galaxies, they appear to become more flattened as they become fainter. When dividing the total sample of dwarf galaxies at a $M_V=-7.5$ threshold\footnote{This somewhat arbitrary threshold was used as it divides the complete MW dwarf galaxy sample in two subsamples of similar sizes (10 and 11 members). This subsampling, that we do not claim to be physical in any way, also separates galaxies without any clear evidence of complex stellar populations from those which do, like LeoT \citep{irwin07,dejong08b} and CVnI \citep{martin08a}.}, more than a third of the faint systems (4 out of 11) have ellipticities that are higher than 0.5. On the other hand, this is the case for only a single bright satellite (UMi; or 2 if one accounts for Sgr). Furthermore, three of these highly flattened faint systems (Her, UMaI, UMaII) are reasonably well populated systems in the SDSS ($N_*=128$, 70 and 296 stars respectively) and with only small uncertainties on $\epsilon$.

To quantify this difference in the ellipticity, the right panel of Figure~\ref{ell_fig} presents the probability distribution functions of the faint (full line) and bright subsamples (dashed line). The arrows correspond to the mean of the two distributions, that are significantly different (a $4\sigma$ difference): $\epsilon=0.32\pm0.02$ and $\epsilon=0.47\pm0.03$, respectively (the uncertainties are determined from a Monte-Carlo resampling of the subsamples, redrawing the individual $\epsilon$ values following the uncertainties of Table~1). A KS test also yields a very low probability (0.4\%) that the two subsamples are derived from the same underlying distribution. The scatter of the PDFs is, however, hard to constrain as it results from two effects: from the larger uncertainties of faint systems, especially the rounder ones ($\epsilon<0.5$), but also from the presence of objects whose uncertainties span a large range in ellipticities. In contrast, the majority of bright satellites are clustered in the range $\epsilon=0.25-0.35$ (Car, Dra, For, Scu, Sex, LeoT).

\subsection{Why are the faintest satellites more flattened?}

Several possible explanations for these flat dwarf galaxies have already been laid out in \citet{coleman07} for the case of Her, but are worth recalling and developing in a broader perspective here.

\subsubsection{Could the faintest satellites be flattened by rotation?}
Current spectroscopic surveys, albeit usually targeting only a few tens of stars in each satellite, show no obvious sign of rotation. Even if the systems were rotationally supported, they would be approximately oblate. As we can assume we are seeing them at random viewing angles, their intrinsic flattening would be much more than $<\epsilon>=0.47\pm0.03$. In that case, they would have implausibly flat shapes and implausibly low intrinsic velocity dispersions, given their already low measured velocity dispersions in the range $4-12\kms$ \citep{kleyna05,munoz06b,martin07a,simon07}.

\subsubsection{Could the stars in the faintest satellites trace the shape of more elongated dark matter halos?}
It has been shown that dissipative gas infall during galaxy formation can induce changes in the shape of dark matter (DM) halos and make them rounder than in the absence of any baryons (see e.g. \citealt{dubinski94,kazantzidis04,maccio07}). The lower baryon/star content of the recently discovered MW satellites \citep{kleyna05,munoz06b,martin07a,simon07} would permit the dark matter halo to remain triaxial, often nearly prolate, for these systems. Assuming stars trace the DM potential, they would then appear more elongated in the faintest systems.

\citet{kuhlen07} have studied the shape of the DM subhalos in the Via Lactea simulation of a Milky Way size $\Lambda$CDM halo. They follow subhalos down to a mass of $10^8M_\sun$, not much higher than those of the faint MW dwarf galaxies studied here, and find them to have a more elliptical shape in their central regions. This could provide an alternate explanation for the shape of the faintest satellites, which are also the smallest (because the larger ones cannot be found in the SDSS; \citealt{koposov08}): stars probe a smaller region of the subhalo, and therefore trace out a less spherical potential. This latter argument cannot, however, explain the shape of the three satellites with the highest $\epsilon$ (Her, UMaI and UMaII) as all three are large satellites with $r_h>140\pc$.

The main issue with this scenario is the assumption of stars distributing themselves in the DM subhalo so as to follow the shape of underlying mass potential for which no cogent mechanism exists.

\subsubsection{Could the elongation be the sign of tidal distortion?}
The best example of tidally induced elongation in the Local Group is given by the Sagittarius dwarf that is clearly being disrupted by the Milky Way as it leaves behind stellar streams wrapped around the Galaxy \citep{ibata01b,majewski03}. This results in an ellipticity of $\sim0.65$ for the (already disrupting) main body of the dwarf \citep{majewski03}, close to the values that we have measured for Her, UMaII and, to a lesser extent UMaI. The absence of any direct evidence of stellar streams originating from the faint dwarf galaxies could easily be explained by their faintness that implies that, if they exist, such features should be very sparsely populated. UMaII has also been proposed to be the progenitor of the Orphan Stream, which would lend support to this tidal distortion scenario, at least for this satellite.

As explained in \citet{coleman07}, tidal distortion requires the satellite to pass close enough to the MW to be stripped. Simple arguments can be used to estimate the required pericenter of the galaxy's orbit as $R_\mathrm{peri}\simeq r_\mathrm{gal}\cdot\sigma_\mathrm{MW}/\sigma_\mathrm{gal}$ with $r_\mathrm{gal}$ its size scale, $\sigma_\mathrm{gal}$ its velocity dispersion and $\sigma_\mathrm{MW}\simeq150\kms$ that of the Milky Way. For Her, UMaI and UMaII, the inferred pericenter is of the order of $\sim10\kpc$ which makes for very eccentric orbits in the case of UMaI ($R_\mathrm{apo}>97\kpc$) and Her ($R_\mathrm{apo}>132\kpc$), with an eccentricity of $e\gta0.90$ for both. In addition, this latter object is certainly not at apogalacticon since it shows a large receding motion from the MW ($v_r=145\kms$, corrected for the Solar motion). Very eccentric orbits are not unheard of in dark matter simulations of galactic groups. For example, \citet{gill04} show that their surviving satellites at $z=0$ have a mean orbital eccentricity of $<e>=1-R_\mathrm{peri}/R_\mathrm{apo}=0.61$, with a standard deviation of 0.19. \citet{benson05} also finds that half of infalling dark-matter substructures are on near parabolic orbits ($e=1.0\pm0.1$). Moreover, \citet{sales07b} identified in their simulations of $L_*$ galaxies that about a third of their satellite dark subhalos have `unorthodox' (including high eccentricity) orbits from three body interactions in the galactic halo. In this scenario, the satellites most severely stripped by tides would be both the least luminous and the most elongated.

However, \citet{penarrubia08b} have recently argued that the observed size and central radial velocity dispersions of the new SDSS satellites are generally not low enough to be disrupted versions of the Fornax, Draco or Sagittarius dwarfs. So this explanatory scenario also appears problematic.

\medskip

Overall, none of these three scenarios is entirely satisfactory but we believe that the latest, elongation induced by tidal interaction with the Milky Way, may be the least problematic. Indeed, there \emph{is} a known stellar stream that could be produced by the disruption of UMaII, the Orphan Stream, and a few MW satellites with very eccentric orbits is far from impossible. Moreover, the importance of the \citet{penarrubia08b} caveat depends, for each satellite, on the amount of stellar loss, which is all but constrained. The new satellites could therefore correspond to a murky sample of galaxies at different stages of their disruption, strongly hampering any analysis of the population as a whole. 

\section{Conclusion}

We have presented a rigorous and homogeneous (re-)analysis of the structure of the stellar bodies of the recently discovered Milky Way satellites. The application of a Maximum Likelihood algorithm to SDSS data yields robust estimates of the centroid, position angle, ellipticity, exponential half-light radius and number of member stars (within the SDSS) for the 15 faintest candidate dwarf galaxies around the Galaxy. With the assumption of stellar population models, derived through the analysis of the same dataset, we have also studied the impact of `CMD shot-noise' on the observed luminosities and stellar masses of these satellites. The magnitude of the faintest systems can be hampered by a large scatter (of up to $\sim1$\,mag) that is due solely to the individual evolution of their very few stars. Stellar mass estimates suffer less from this sparseness but, obviously, depend strongly on the assumed IMF. Finally, we have shown that except for tentative evidence of deviations from the best model in the CVnI and UMaII dwarf galaxies, the apparently distorted morphology of the satellites in the SDSS can be entirely accounted for by shot-noise. This does not mean that some distortion is not present, but only that deeper datasets are required: one should not rely on a visual inspection of stellar density maps as the structure in these is strongly driven by the low number of stars that are observed in each satellite. 

In our discussion, we focussed on the remarkable result that the faintest satellites are, on average, highly flattened. We put forth three lines of explanations, but none of these three scenarios is entirely satisfactory. However, we believe that tidally induced elongation may be the least problematic and is certainly the easiest to (dis)prove from observed distortions or tidal streams from the stellar maps of these dwarf galaxies. This however requires deeper follow-up data than the SDSS since our search of such distortions yields mainly null-detections when accounting for Poisson noise. Alternatively, a final answer will strongly benefit from deeper systematic surveys such as the PanSTARRS $3\pi$ survey \citep{kaiser06} that should significantly expand the SDSS harvest of faint satellites of the Local Group.

\acknowledgments
N. F. M. gratefully acknowledges Sebastian Jester, Eric Bell and Knud Jahnke for statistically relevant discussions as well as Andrea Macci\`o for cosmologically motivated conversations. It is also a pleasure to thank Matthew Coleman and Alan McConnachie for helpful comments on this work. Finally, we would like to acknowledge the referee for her/his thorough reading of the manuscript.

Funding for the Sloan Digital Sky Survey (SDSS) and SDSS-II has been provided by the Alfred P. Sloan Foundation, the Participating Institutions, the National Science Foundation, the U.S. Department of Energy, the National Aeronautics and Space Administration, the Japanese Monbukagakusho, and the Max Planck Society, and the Higher Education Funding Council for England. The SDSS Web site is http://www.sdss.org/.

The SDSS is managed by the Astrophysical Research Consortium (ARC) for the Participating Institutions. The Participating Institutions are the American Museum of Natural History, Astrophysical Institute Potsdam, University of Basel, University of Cambridge, Case Western Reserve University, The University of Chicago, Drexel University, Fermilab, the Institute for Advanced Study, the Japan Participation Group, The Johns Hopkins University, the Joint Institute for Nuclear Astrophysics, the Kavli Institute for Particle Astrophysics and Cosmology, the Korean Scientist Group, the Chinese Academy of Sciences (LAMOST), Los Alamos National Laboratory, the Max-Planck-Institute for Astronomy (MPIA), the Max-Planck-Institute for Astrophysics (MPA), New Mexico State University, Ohio State University, University of Pittsburgh, University of Portsmouth, Princeton University, the United States Naval Observatory, and the University of Washington.

{\it Facilities:} \facility{SDSS}.

\appendix
\section{Two-dimensional relative likelihood distributions}
\label{appendix}
The following contour plots show, for each satellite, the evolution of the relative likelihood for all the combinations of two parameters that are fit by the ML algorithm. In each panel, outward from the best model that is represented by the dot, contours represent drops of 50\%, 90\% and 99\% of the likelihood with respect to the best model.

\begin{figure*}
\includegraphics[angle=270,width=0.49\hsize]{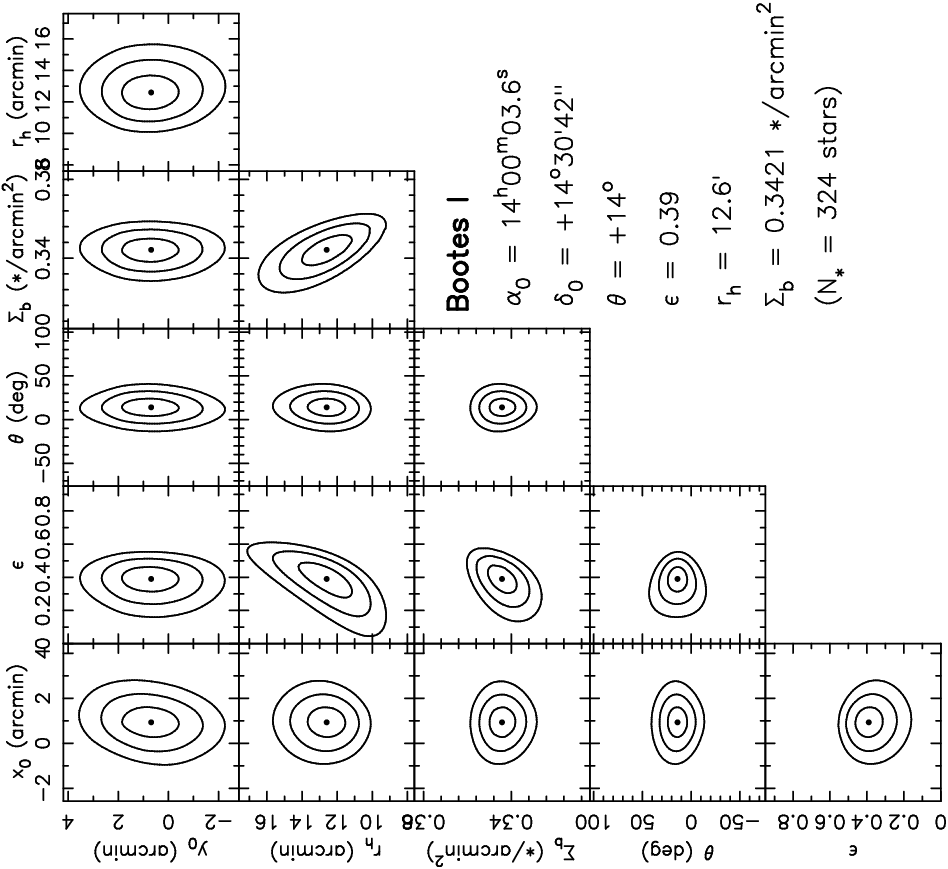}
\includegraphics[angle=270,width=0.49\hsize]{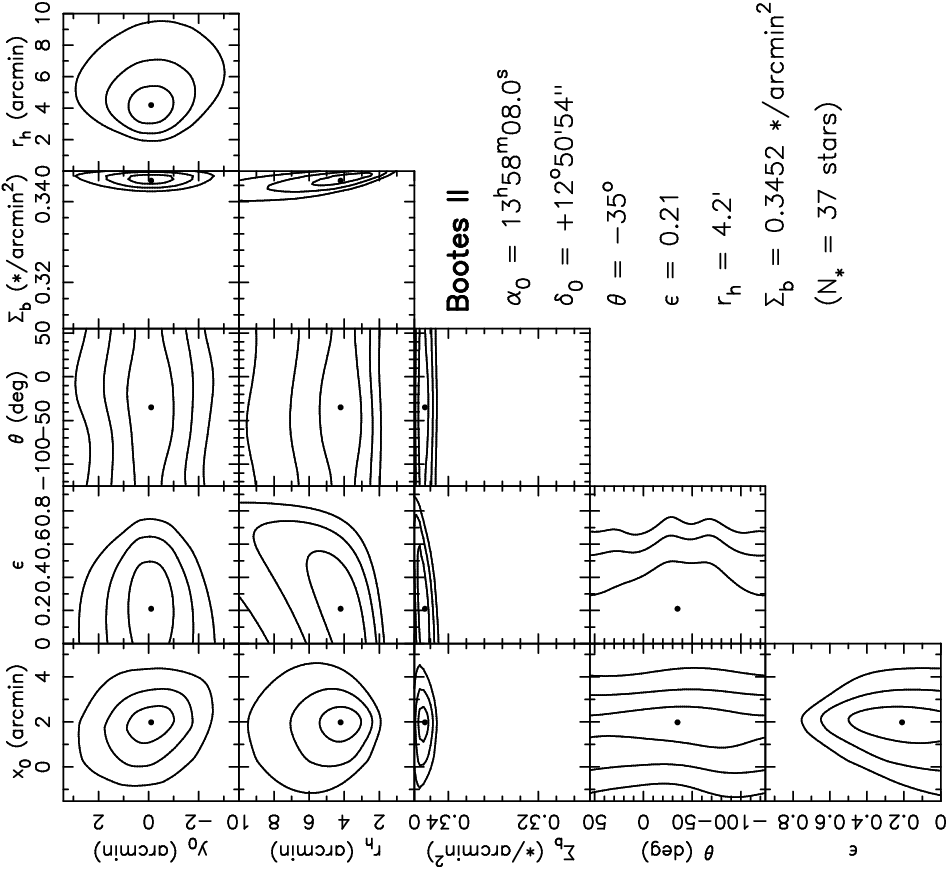}
\includegraphics[angle=270,width=0.49\hsize]{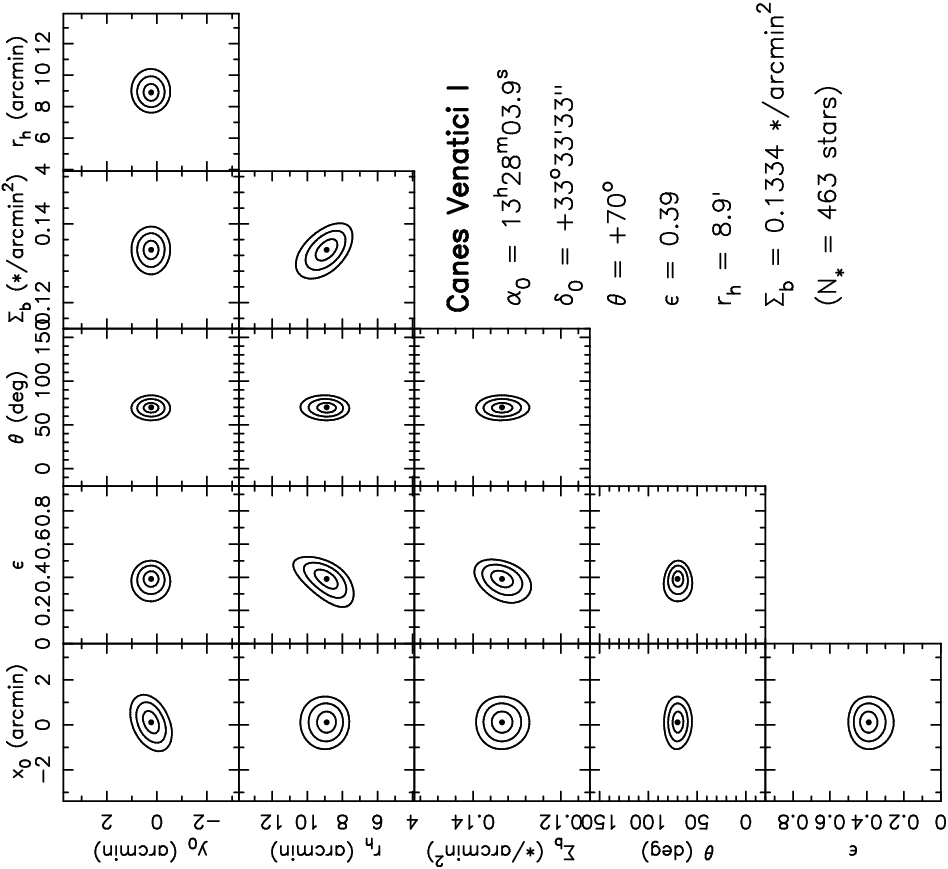}
\includegraphics[angle=270,width=0.49\hsize]{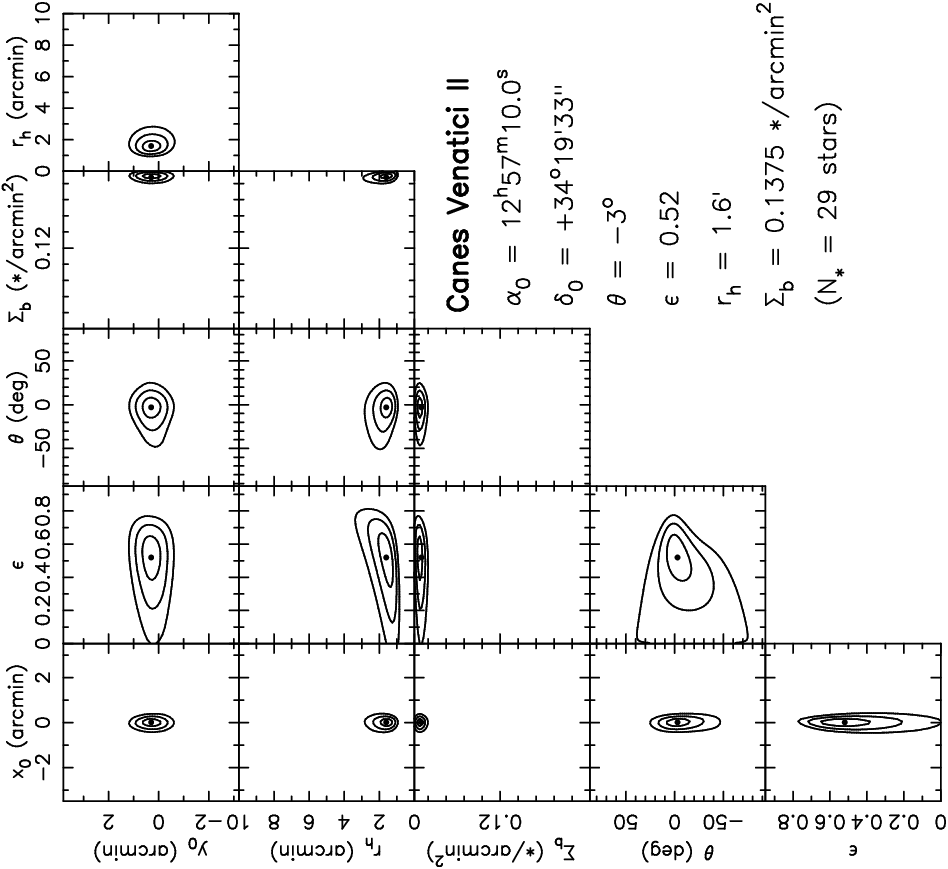}
\caption{\label{ML_plots} Evolution of the relative likelihood for all the combinations of two parameters that are fit by the ML algorithm for BooI, BooII, CVnI and CVnII.}
\end{figure*}

\begin{figure*}
\includegraphics[angle=270,width=0.49\hsize]{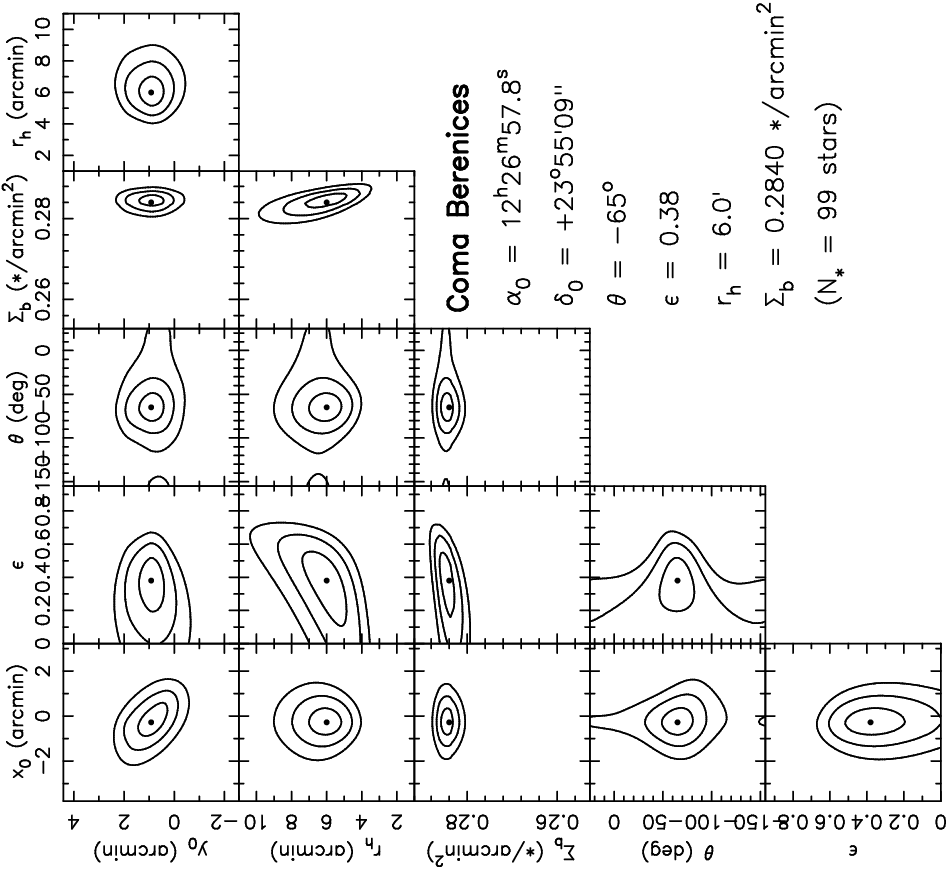}
\includegraphics[angle=270,width=0.49\hsize]{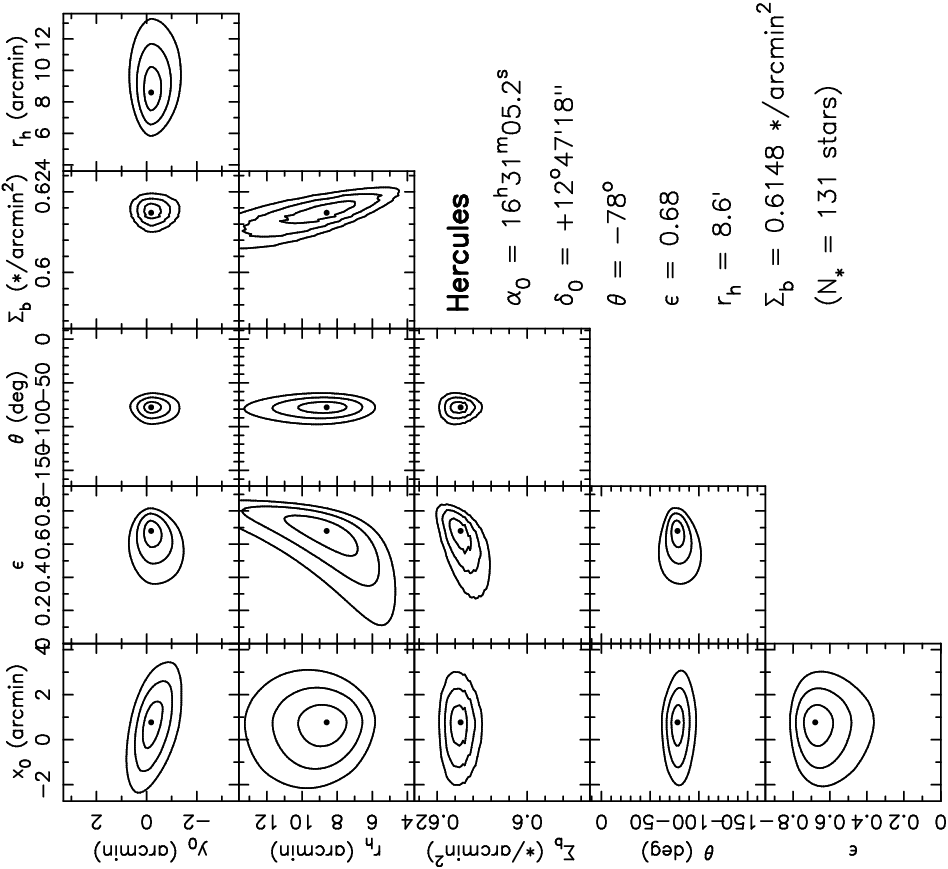}
\includegraphics[angle=270,width=0.49\hsize]{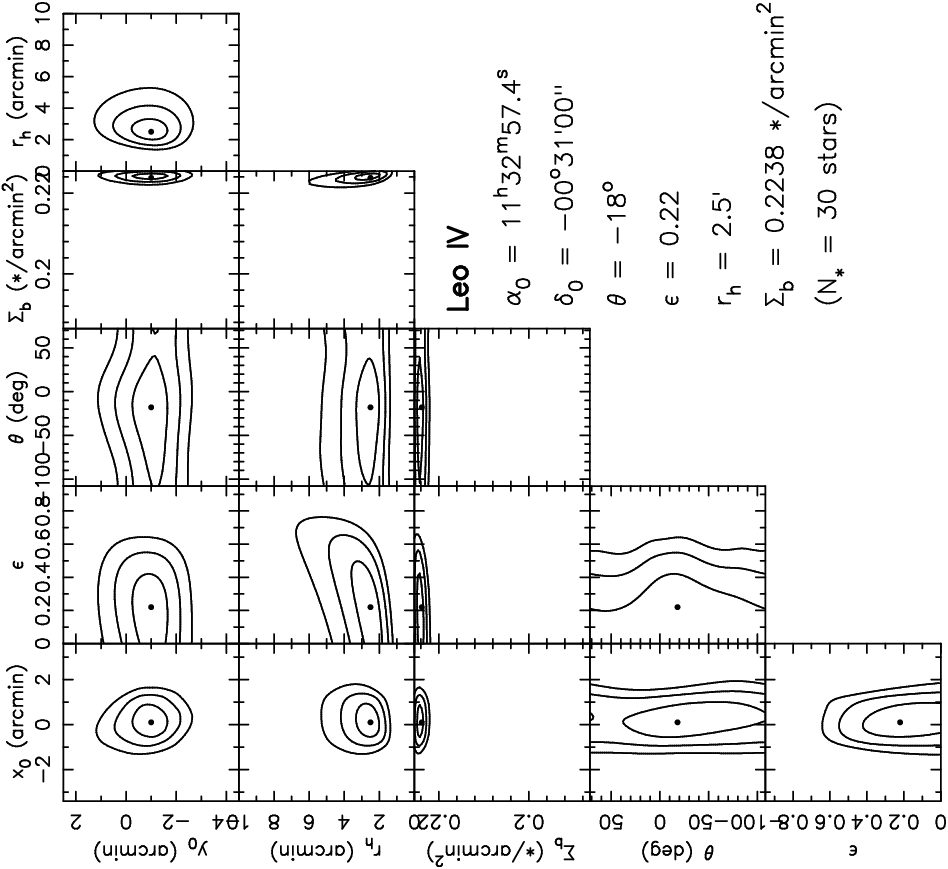}
\includegraphics[angle=270,width=0.49\hsize]{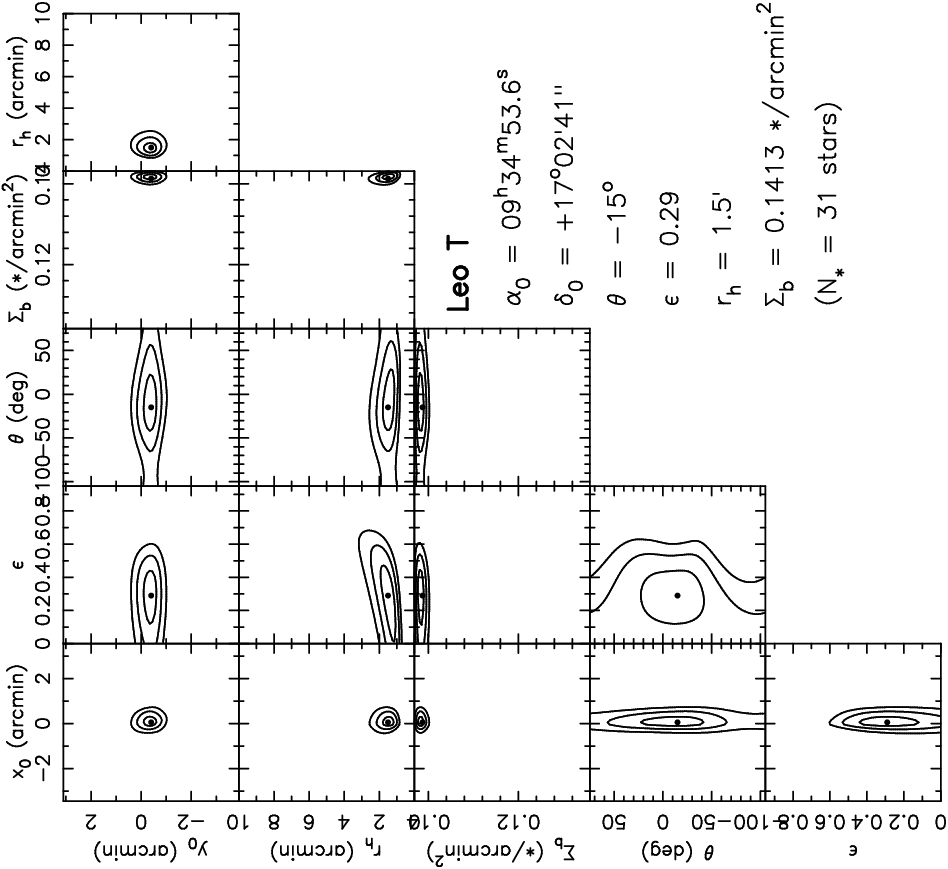}
\caption{Same as Figure~\ref{ML_plots} for Com, Her, LeoIV and LeoT.}
\end{figure*}

\begin{figure*}
\includegraphics[angle=270,width=0.49\hsize]{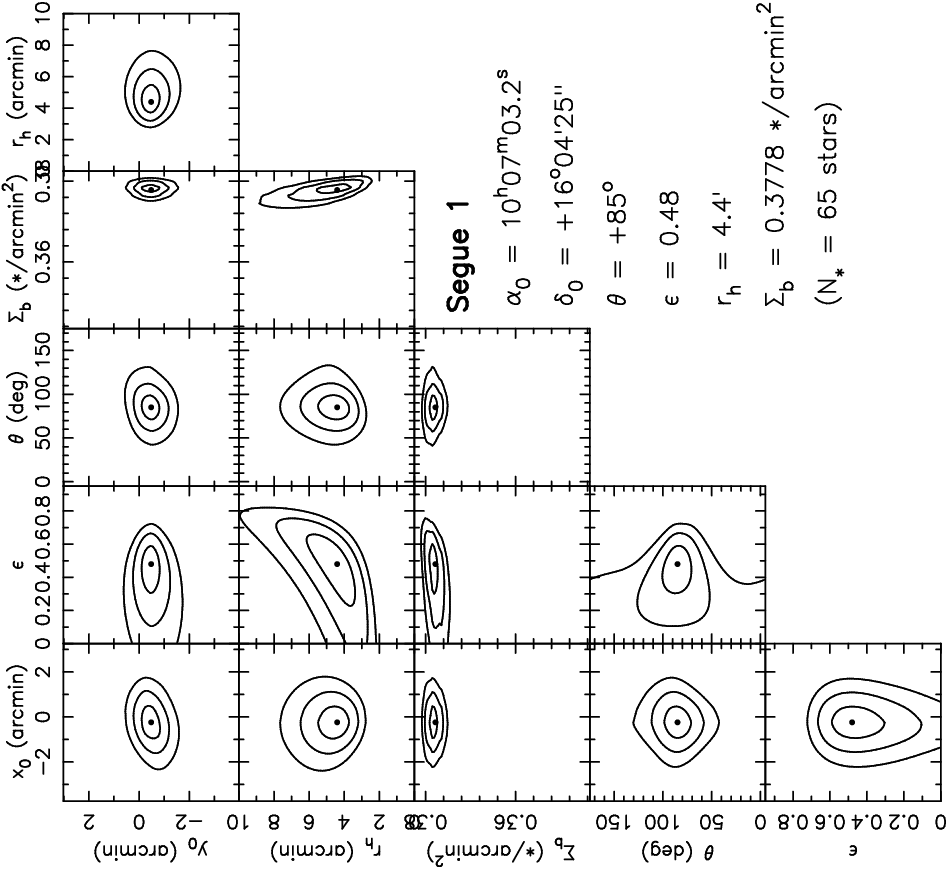}
\includegraphics[angle=270,width=0.49\hsize]{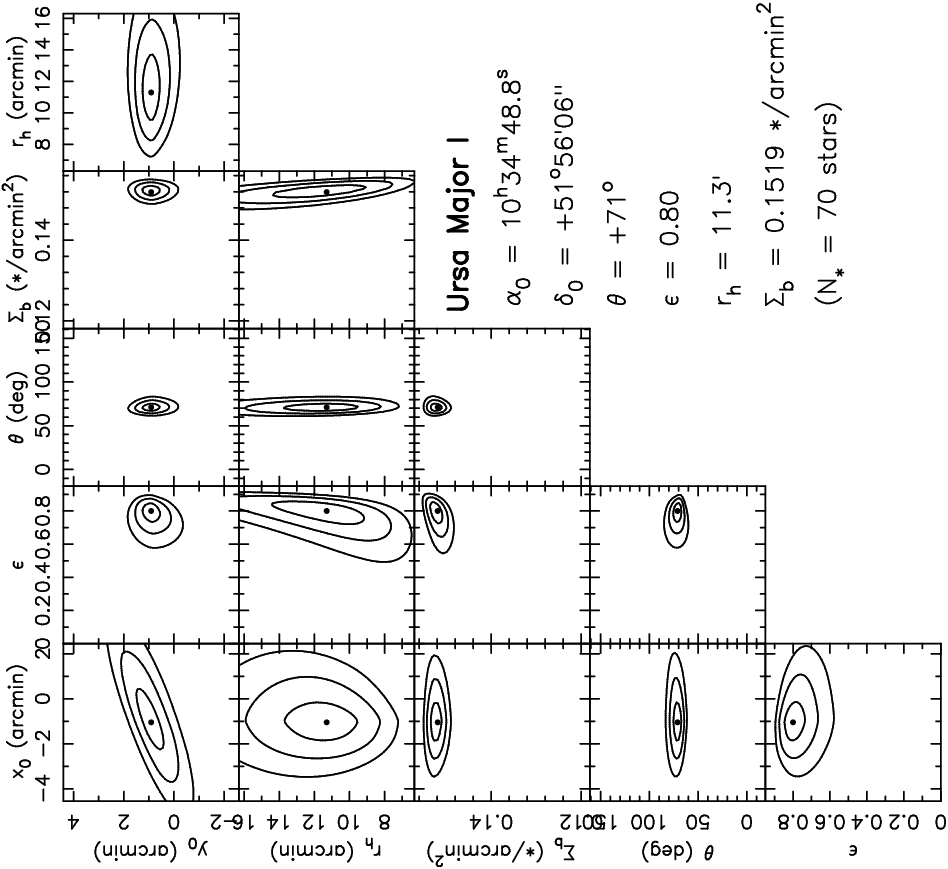}
\includegraphics[angle=270,width=0.49\hsize]{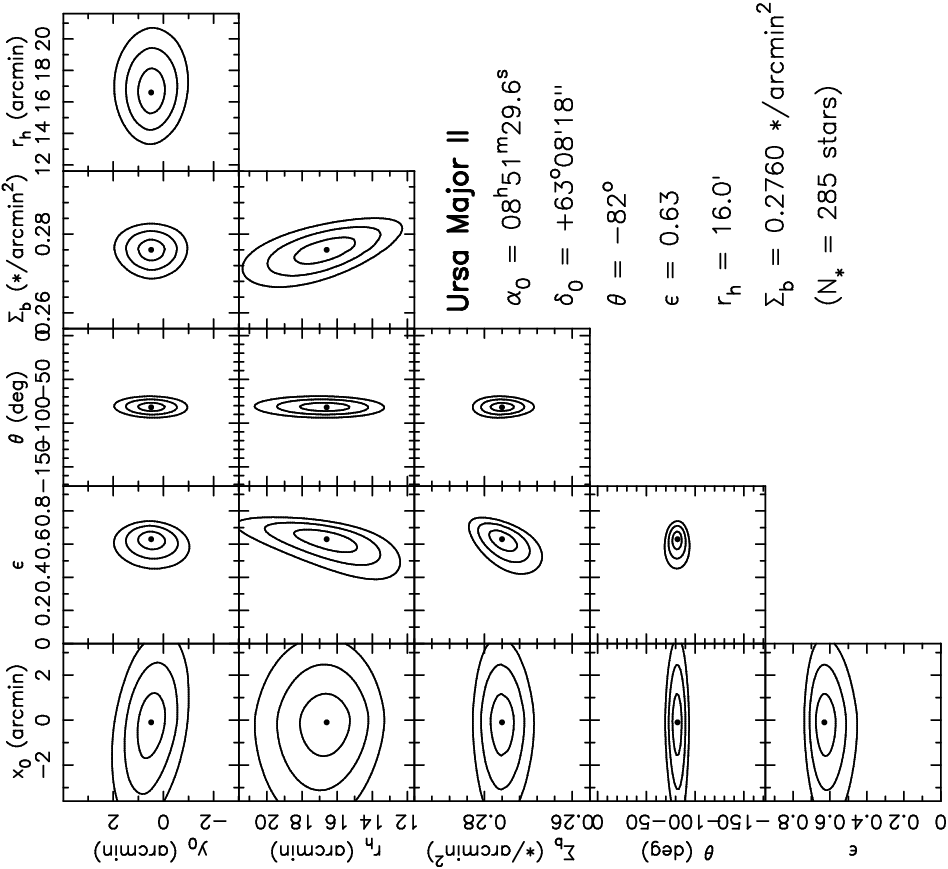}
\includegraphics[angle=270,width=0.49\hsize]{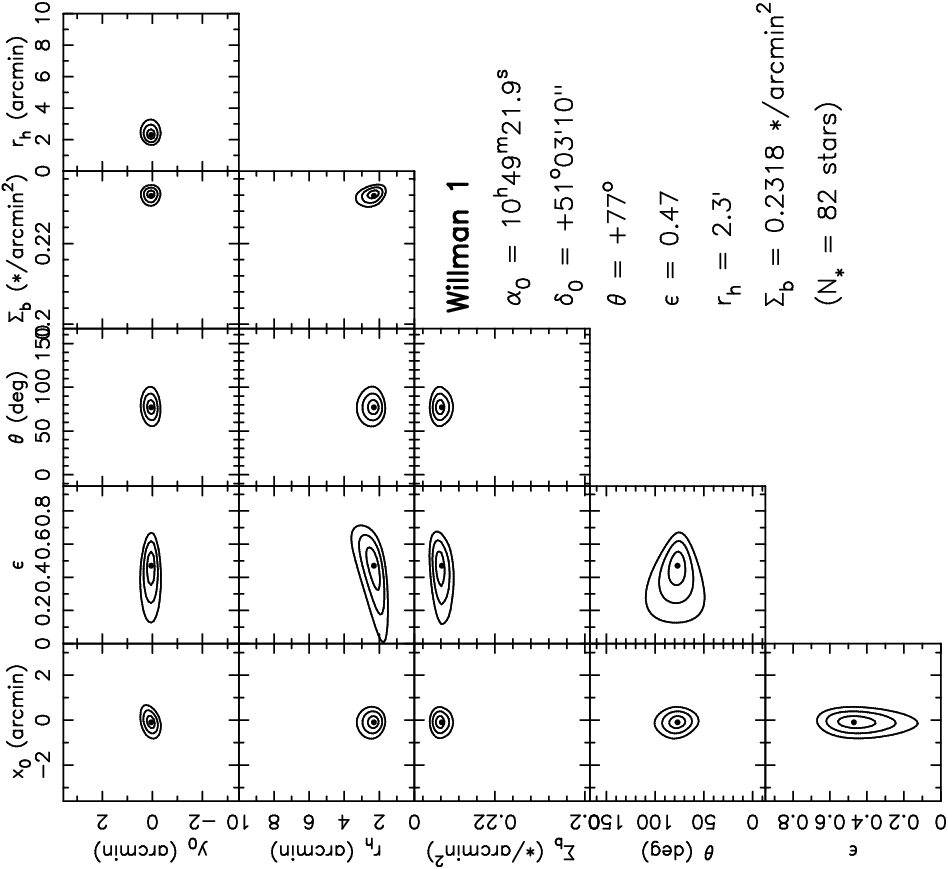}
\caption{Same as Figure~\ref{ML_plots} for Seg1, UMaI, UMaII and Wil1.}
\end{figure*}

\begin{figure*}
\includegraphics[angle=270,width=0.49\hsize]{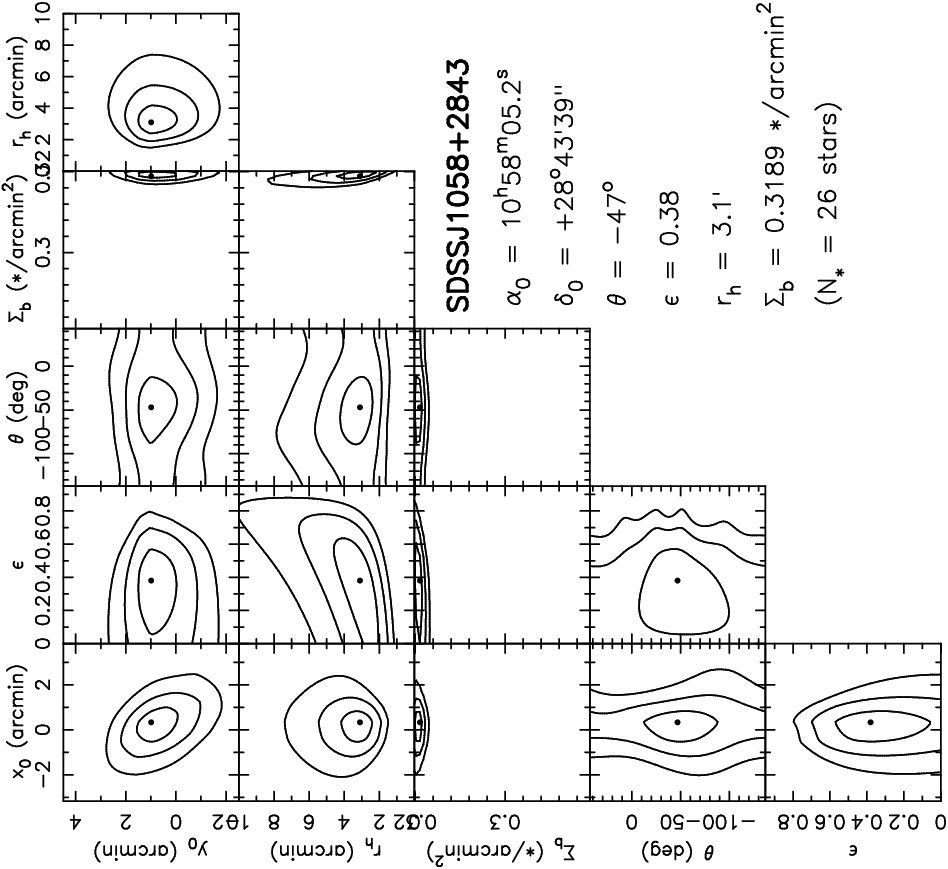}
\includegraphics[angle=270,width=0.49\hsize]{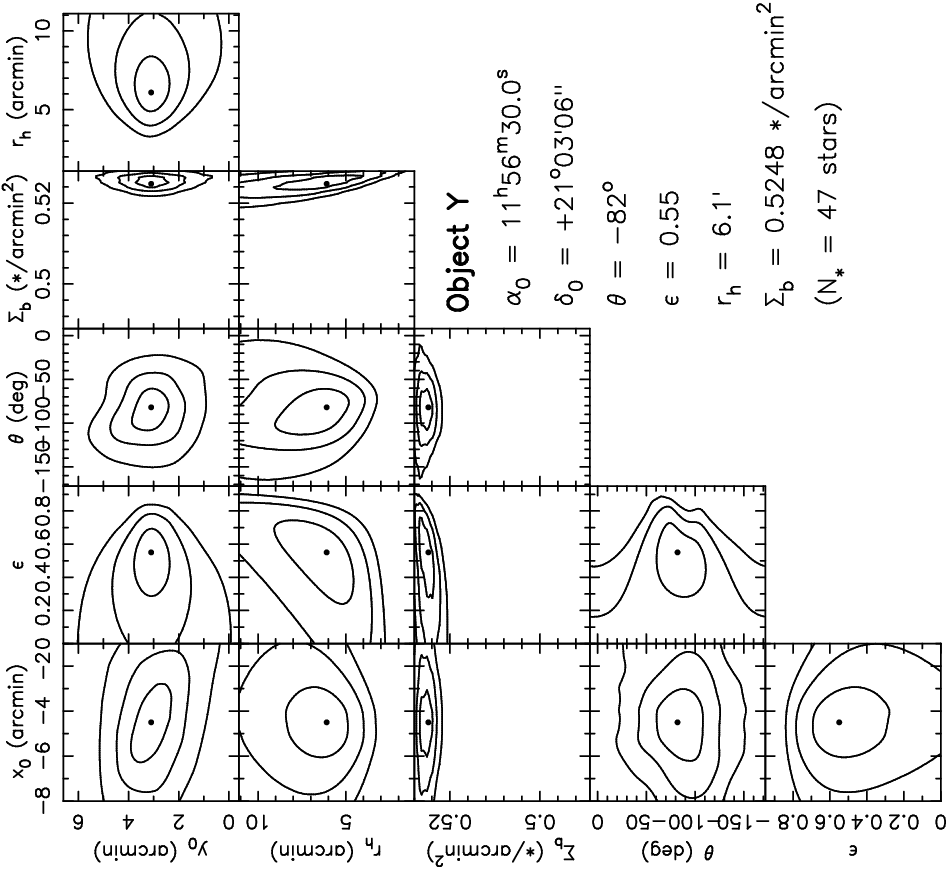}
\includegraphics[angle=270,width=0.49\hsize]{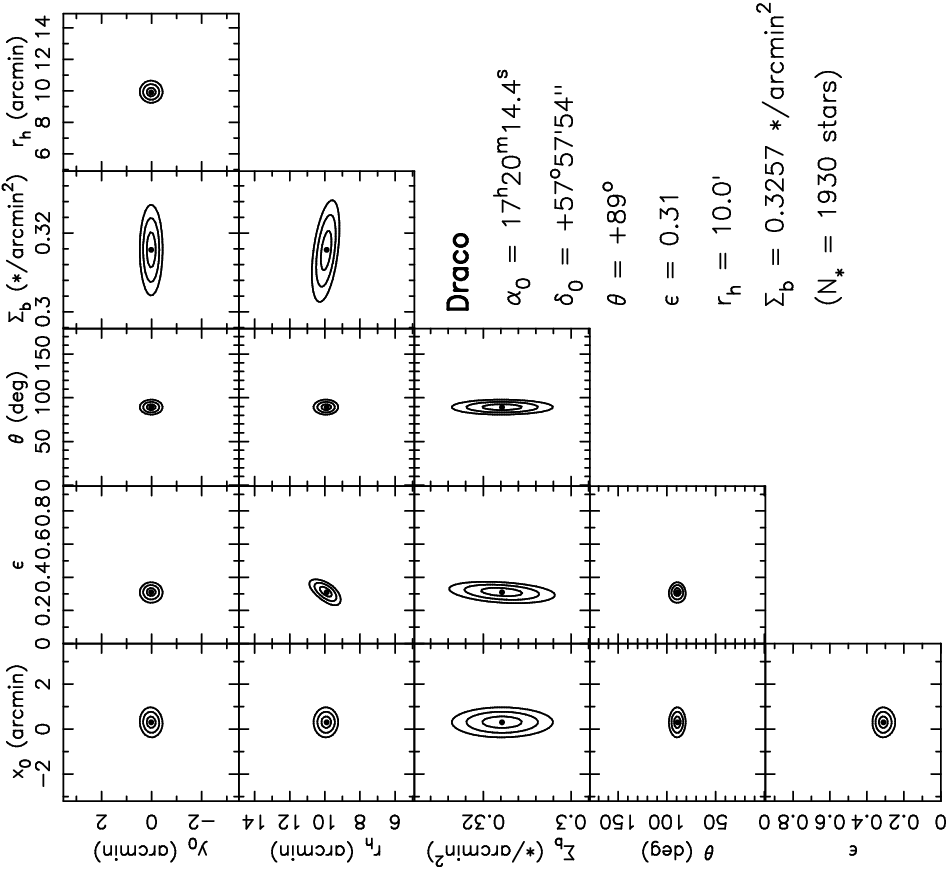}
\caption{Same as Figure~\ref{ML_plots} for SDSSJ1058+2843, Object~Y and Dra.}
\end{figure*}

\clearpage

\clearpage

\end{document}